%% file: main.tex
\documentclass[onefignum,onetabnum]{siamonline171218}

\input{preamble}

\input{ex_shared}

\ifpdf
\hypersetup{
  pdftitle={Mixture},
  pdfauthor={Jaimungal, Pesenti}
}
\fi

\begin{document}

\maketitle
\begin{abstract}
We consider the problem where an agent aims to combine the views and insights of different experts' models. Specifically, each expert proposes a diffusion process over a finite time horizon. The agent then combines the experts' models by minimising the weighted Kullback--Leibler divergence to each of the experts' models. We show existence and uniqueness of the barycentre model and prove an explicit representation of the Radon--Nikodym derivative relative to the average drift model. We further allow the agent to include their own constraints, resulting in an optimal model that can be seen as a distortion of the experts' barycentre model to incorporate the agent's constraints. 
We propose two deep learning algorithms to approximate the optimal drift of the combined model, allowing for efficient simulations. The first algorithm aims at learning the optimal drift by matching the change of measure, whereas the second algorithm leverages the notion of elicitability to directly estimate the  value function. The paper concludes with an extended application to combine implied volatility smile models that were estimated on different datasets. 
\end{abstract}

\begin{keywords}
  Mixtures of Experts, Kullback--Leibler, Barycentre, Model Combination, Ensemble Model, Deep learning, Elicitability, Volatility Smiles
\end{keywords}

  

\section{Introduction}
In many situations a modeller or agent aims to find a single model to, e.g., predict, forecast, classify, or approximate observed patterns. While model selection procedures provide rankings of the different models' performances --- thus allowing the agent to select one model --- an alternative to disregarding all but one model is model combination. In the statistical literature, so-called forecast ensemble or superensemble models have been widely applied to weather prediction as they are known to improve deterministic forecasts, see e.g., \cite{Gneiting2011JASA} and reference therein. A variant of a statistical superensemble that allows for predictive distribution functions is to regress a quantity of interest (e.g., air pressure, temperature) onto forecasts stemming from different models \cite{Gneiting2011JASA}. Averaging different predictions has also been used in machine learning, where early examples include bagging predictors \cite{Breiman1996ML}, Adaptive, Reweighting and Combining (ARCing) algorithms  and Adaboost \cite{Breiman1999NC}. To improve the architecture of neural networks, \cite{Hansen1990IEEE} propose neural networks ensembles (averaging different neural network architectures), which have been applied in, e.g., credit scoring \cite{West2005neural} and oil price forecasting \cite{Yu2008forecasting}. We also refer to \cite{Liu2024WP} for an application to fusion of generative models.

This work focuses on model combinations of dynamic stochastic models in continuous-time. A related stream of literature, though in discrete-time, is called dynamic time warping (see e.g., \cite{berndt1994using}), which aims at matching or detecting similarities in time series data by finding a ``warping path'' that minimises a distance between the time series. Applications of dynamic time warping are widespread and range from classification of genome signals \cite{Skutkova2013classification}, to speech recognition \cite{Juang1984hidden}, and clustering of financial stocks \cite{durso2021AOR}. 

Differently from dynamic time warping, which is a data driven approach based on time series sample paths, we propose to average different dynamics of stochastic processes. Furthermore, while most model averaging methodologies optimise for the weights to be associated with each model, in our setting the weights are given. Thus, a key contribution of this work is to characterise the dynamics of the stochastic process that minimises a weighted distance to each model, that is finding the stochastic barycentre model.

In this manuscript, we consider a finite number of experts, each having their own belief about the dynamics of a continuous-time diffusion process. The experts' models are characterised by different probability measures, under which the process follows the experts' dynamics. An agent then aims at combining the different experts' models by finding the probability measure that minimises the weighted Kullback--Leibler (KL) divergence to all models. The resulting dynamics of the stochastic process is described by the barycentre model. Moreover, we allow the agent to include their own views, described by expectations of functions of the terminal values of the process or expected running cost, which results in a constrained barycentre model. Key contributions of this work are as follows. First, we derive the optimal value function of our optimisation problem and a succinct representation of the Radon--Nikodym (RN) derivative of the optimal probability measure to, what we term, the average drift measure. We find that the optimal RN derivative is of Escher type albeit different to the classical solution of maximum entropy, see e.g., \cite{Csiszar1975AP}. Second, we prove that our constraint optimisation problem is equivalent to an optimisation problem of distorting the expert's barycentre model to include the agent's constraints. The latter optimisation problem, that is distorting stochastic processes to include constraints has been studied in \cite{Kroell2024IME} and \cite{Jaimungal2024SICON}. Third, we propose two deep learning algorithms to approximate the optimal probability measure and thus the dynamics of the constrained barycentre process, allowing for simulation under the constrained barycentre model. Fourth, we apply our framework to combining implied volatility smile models that were estimated on different sets of real data.

Here, we utilise the KL divergence to quantify divergences between probability measures on the path space of stochastic processes. Alternatives could be distances on the space of probability measures stemming from optimal transport theory, such as the Wasserstein distance. There are, however, two caveats: first, calculating a Wasserstein barycentre between random variables is in general an NP hard problem \cite{Altschuler2022SIAM-MDS}, highlighting the curse of dimensionality associated with the Wasserstein distance; second, as we work in continuous time and over a finite time horizon, the Wasserstein distance needs to be replaced with the adapted (also called bicausal) Wasserstein distance. The adapted Wasserstein distance between stochastic processes is only known in special cases \cite{bion2019wasserstein,Backhoff2022WP, gunasingam2025adapted, robinson2024bicausal}, and most recent results on barycentre with adapted Wasserstein barycentre pertain to existence and uniqueness \cite{Bartl2021WP,Acciaio2024WP-bary}, while with the KL divergence we obtain a closed-form solution of the optimal RN derivative for arbitrary dimension of the stochastic process. 

The manuscript is structured as follows. \Cref{sec:main} introduces the experts' models and the agent's optimisation problem \eqref{opt:P}. \Cref{sec:barycentre} establishes the solution to the pure barycentre. In \Cref{sub-sec:bary-two} we consider the special case of two expert models and in \Cref{sub-sec:bary-OU} the special case of Ornstein-Uhlenbeck processes. The solution to the KL barycentre with beliefs is detailed in \Cref{sec:solution}. Key results are the concise representation of the optimal change of measure (\Cref{prop:RN-representation}) and the recasting of optimisation problem \eqref{opt:P} into an optimisation problem of minimally distorting the barycentre measure to include the agent's constraints (\Cref{prop:equivalent-opt}). In \Cref{sec:algos}, we propose two deep learning algorithms to approximate the drift of the optimally combined model. \Cref{sec:sim-ex} illustrates and compares these algorithms on simulated examples. A financial application to combining implied volatility smiles models, that were estimated on different datasets, is provided in \Cref{sec:IV-smile}.

\section{Constrained Kullback--Leibler barycentre}
\label{sec:main}
This section first introduces the experts' models and the optimisation problem --- minimising the weighted KL divergence between the experts' models with constraints --- that the agent aims to tackle. Second, using a stochastic control approach, we solve the optimisation problem and characterise the optimal model. Third, we relate the optimisation problem to a modified barycentre model, in the spirit of \cite{Jaimungal2024SICON}. 

\subsection{Experts' opinions and agent's problem}
We work on a complete and filtered probability space, denoted by $\left(\Omega, \mathbb{P}, \F,(\F_t)_{t\in\T}\right)$, satisfying the usual conditions of right continuity, on which we have a $d$-dimensional Brownian motion $B=(B_t)_{t\in\T}$ (with independent components) and the filtration is the natural one generated by $B$. We further have a $d$-dimensional process $X=(X_t)_{t\in\T}$, that the agent is interested in modelling, and a set of experts $\mcK:=\{1,\dots,K\}$. Each expert $k\in\mcK$ has a probability measure $\P^{(k)}$ --- called the ``model'' of expert $k$ --- which is equivalent to $\P$ ($\P\sim\P^{(k)}$), and believes that under $\P^{(k)}$ the process $X$ satisfies the stochastic differential equation (SDE)
\begin{equation}\label{eq:SDE-X}
    \d X_t = \mu^{(k)}(t,X_t)\,\dt + \sigma(t,X_t)\,\dW_t^{(k)}\,,
\end{equation}
where $W^{(k)}=(W^{(k)}_t)_{t\in\T}$ denotes a $d$-dimensional $\P^{(k)}$-Brownian motion, $\mu^{(k)}(t,x):\R_+\times \R^d\to \R^d$ the drift, and $\sigma(t,x):\R_+\times \R^d \to \S^d_{++}$, where $\S^d_{++}$ is the set of $d$-dimensional matrices such that (s.t.) $\sigma^\intercal\sigma$ is strictly positive definite, the volatility. That is, each expert has a different view on the drift of the process $X$. If the experts had differing views on the volatility, so that expert--$k$ wanted a diffusion coefficient of $\sigma^{(k)}(t,X_t)$, then  the KL divergence between the experts' models is infinite, and hence we could not meaningfully determine a weighted KL barycentre model. Instead, one could regularise the problem, by e.g., introducing a new process $Y=(Y_t)_{t\in\T}$ and a new Brownian motion $B^Y$ (independent of all others) that mean-reverts to $\sigma^{(k)}(t,X_t)$. Then the expert model could be defined as
\begin{subequations}
\begin{align}
    \d X_t = \mu^{(k)}(t,X_t)\,\dt + Y_t\,\dW_t^{(k)}\,,
    \\
    \d Y_t = \tfrac1\ep \big(\sigma^{(k)}(t,X_t)-Y_t\big)\,\dt + \dW_t^{Y(k)},
\end{align}%
\end{subequations}%
where $W^{Y(k)}$ is a $\P^{(k)}$--Brownian motion independent of the other Brownian motions $W^{(k)}$. As $\ep\downarrow0$, the process $Y$ is pinned to $\sigma^{(k)}(t,X_t)$. In this manner, the setting where experts have differing volatilities may be cast into our original setting by increasing the state space dimension by $1$. 

Returning now to our original setting, the next set of assumptions guarantee that a strong solution to \eqref{eq:SDE-X} exists.
\begin{assumption}\label{asm:strong=sol-SDE}
The functions $\mu^{(k)}(\cdot,\cdot)$, for all $k\in\mcK$, and $\sigma(\cdot,\cdot)$ satisfy the linear growth and Lipschitz continuity conditions. That is here exists a constant $C_1 < \infty$, such that for all $\tT$ and all $x \in \R^d$, 
\[
    \|\sigma(t, x)\|^2 + |\mu^{(k)}(t,x)|^2 \le C_1 (1 + |x|^2)\,,    
\]
where $|\cdot|$ denotes the Euclidean norm and $||\cdot||$ the Frobenius norm. Moreover, there exists a constant $C_2< \infty$, such that for all $\tT$ and all $x, y \in \R^d$,
\[
    \|\sigma(t,x)- \sigma(t,y)\|^2 + |\mu^{(k)}(t,x) - \mu^{(k)}(t,y)|^2 \le C_2\, |x-y|^2\,.
\]
\end{assumption}

With a slight abuse of notation, when there is no confusion, we use the notation $\mu^{(k)}_t:=\mu^{(k)}(t,X_t)$ and $\sigma_t:=\sigma(t,X_t)$.

While the experts have differing views --- in particular they disagree on the drift of $X$ --- an agent wishes to combine the opinions of these experts and assigns each expert a weight $\pi=(\pi_1,\dots,\pi_K)$ with $\pi_k\in[0,1]$, $k\in\mcK$, and $\sum_{k\in\mcK}\pi_k=1$. That is, the agent  aims at reflecting the experts' opinions and incorporating them with their own belief into a combined probability measure $\Q$, which we call the combined model. The agent's beliefs are specified as follows: for functions $g\colon \R_+\times\R^d \to \R $ and $ f\colon \R^d \to \R $ the agent wishes to ensure that under a probability measure $\Q$ their beliefs 
\[
\E^\Q[f(X_T)]=0\qquad \text{and} \qquad 
\E^\Q\left[\textstyle\int_0^Tg(u,X_u) \du\right] = 0,
\] 
are satisfied.  
Here, and in the sequel, we use the notation $\E^{\Q}[\cdot]$ to denote expectation under the agent's combined model. If $X_T$ is univariate and continuously distributed, the choice $f(x)=\Id_{\{x<q\}} - \alpha$, $\alpha\in(0,1)$, for example, corresponds to the agent requiring that the Value-at-Risk (VaR) at level $\alpha$ of the combined model equals to $q$. The weights $\pi$ could be, e.g., obtained by computing the posterior probability that the data the agent is using stems from expert-$k$'s model.

We propose that the agent combines the models by finding the (weighted) Kullback--Leibler (KL) barycentre of the expert models subject to the agent's constraints. 
Before presenting the formal optimisation problem, we need some conditions on the set of experts' probability measures that the agent aims to combine. The following is a standing assumption in the paper.
\begin{assumption}[Compatibility of Expert Models]
\label{asm:compatibility}
For every $k\in\mcK$,
\begin{equation}
    \E^{\P^{(k)}}\left[e^{\frac12\int_0^T|\bgamma^{(k)}_t |^2\dt}\right] < \infty,
\end{equation}
where $\E^{\P^{(k)}}[\cdot]$ denotes expectation under the $k^{\text{th}}$ expert's model, $\bgamma^{(k)}_t := \sigma_t^{-1}(\mu_t^{(k)}-\bmu_t)$, and $\bmu_t:=\sum_{k\in\mcK}\pi_k\mu_t^{(k)}$ --- the average drift of the experts' models.
\end{assumption}

The average drift of the experts' models $\bmu_t$ characterises the probability measure $\Q[\bmu]$, which we term the ``average drift measure'' and which plays a central role in the exposition. 
\allowdisplaybreaks
Indeed by Theorem  1 in \cite{novikov1973identity}, the conditions in \Cref{asm:compatibility} imply that, for all $k\in\mcK$, $\E^{\P^{(k)}}\left[\frac{\dQ[\bmu]}{\dP^{(k)}}\right]=1$, where the probability measure $\Q[\bmu]$ is defined via the Radon-Nikodym derivative
\begin{equation}
\label{eqn:dQbmu-dPk}
    \frac{\dQ[\bmu]}{\dP^{(k)}} := e^{-\frac12\int_0^T|\bgamma_t^{(k)}|^2\dt
    -\int_0^T\bgamma^{(k)\intercal} \dW^{(k)}_t}.
\end{equation}
Moreover, by Girsanov's Theorem, $\bW:=\left(W_t^{(k)}+\int_0^t\bgamma^{(k)}_u\du\right)_{\tT}$ is a $\Q[\bmu]$--Brownian motion, for every $k\in\mcK$. We also make use of the average drift function $\bmu(t,x):=\displaystyle\sum_{k\in\mcK}\pi_k\,\mu^{(k)}(t,x)$.

We are now in a position to state the agent's optimisation problem:
\begin{align}
    \tag{P}
    \label{opt:P}
    \inf_{\Q\in\mcQ} \sum_{k\in K}
    \pi_k\,\mcH\left[\Q\,\big\| \,\P^{(k)}\right], 
    \qquad s.t.\quad 
    &\E^\Q\left[\int_0^T g(u, X_u)\, \du\right] = 0\, , \quad \text{and} \quad
    \\[0.5em]
    & 
    \notag
    \E^\Q[f(X_T)]=0\,,
\end{align}
where  
\[
\mcH\left[\Q~\|~\P^{(k)}\right]:=\E^{\P^{(k)}}\left[\frac{\dQ}{\dP^{(k)}}\log \frac{\dQ}{\dP^{(k)}}\right]
\]
denotes the KL divergence of $\Q$ relative to $\P^{(k)}$ and $\mcQ$ is the set of probability measures parametrised as
\begin{align*}
    \mcQ:= \Bigg\{\; \Q[\theta]~\Big|~
     & 
     \frac{\dQ[\theta]}{\dQ[\bmu]} = \exp\left\{-\tfrac{1}{2}\int_0^T |\blambda_t|^2\dt-\int_0^T \blambda^\intercal_t\,\dbW_t\right\}\,,
     \\[0.5em]
     &
      \text{where} \quad 
      \blambda_t:=\sigma^{-1}_t\big(\bmu_t-\theta_t\big)
      \,,
    \\[0.5em]
    & \quad \theta:=(\theta_t)_{\tT} \text{ is an $\F$-adapted process,}
    \\[0.5em]
    & 
    \qquad \E^{\Q[\bmu]}\left[\exp\left\{
        \tfrac12 \int_0^T |\blambda_t|^2\,  \dt
        \right\}\right] < \infty,
        \text{ and }
        \E^{\Q[\theta]}\left[\sup_{t\in[0,T]} |X_t|^2\right]<\infty
     \Bigg\}\,.
\end{align*}%
By Theorem 1 in \cite{novikov1973identity}, the condition $\E^{\Q[\bmu]}\left[e^{
        \frac12 \int_0^T |\blambda_t|^2\,  \dt}\right] < \infty$ in the set $\mcQ$ ensures that $\E^{\Q[\bmu]}\left[\frac{\dQ[\theta]}{\dQ[\bmu]}\right]=1$.

Figure \ref{fig:diagram-P} provides a visualisation of problem \eqref{opt:P}. The left panel shows that we first change to the mean drift measure $\Q[\bmu]$, then to a measure $\Q[\theta]$ with arbitrary drift. The middle panel shows that we measure the KL divergence between the experts' measures $(\P^{(k)})_{k\in\mcK}$ and $\Q[\theta]$, and the right panel shows that we aim to have this measure be within the set of expectation constraints $\C$ in problem \eqref{opt:P}.
\begin{figure}
\begin{minipage}{0.28\textwidth}
\centering
\begin{tikzpicture}[    mid arrow/.style={
        postaction={decorate,decoration={
            markings,
            mark=at position .8 with {\arrow{Latex[length=6.4pt, sep=-3.2pt]}}
    }}
  },]

\node[measure] (P1) at (0.5,-1.5) {};
\node[] () at (0.25,-1.5) {${\mathcolor{blue}{\P_1}}$};
\node[measure] (P2) at (-2,0) {};
\node[] () at (-2.35,0) {${\mathcolor{blue}{\P_2}}$};
\node[measure] (P3) at (-1.5,1.25) {};
\node[] () at (-1.85,1.25) {${\mathcolor{blue}{\P_3}}$};
\node[measure] (P4) at (1,0.75) {};
\node[] () at (0.65,1) {${\mathcolor{blue}{\P_4}}$};

\node[measure,fill=red!10] (Qbar) at (-1,0.75) {};
\node[] () at (-1.5,0.75) {${\mathcolor{purple}{\Q[\overline\mu]}}$};
\draw[mid arrow,dashed,color=blue] (P1) to [out=90,in=-25] (Qbar);
\draw[mid arrow,dashed,color=blue] (P2) to [out=-25,in=245] (Qbar);
\draw[mid arrow,dashed,color=blue] (P3) to [out=-25,in=90] (Qbar);
\draw[mid arrow,dashed,color=blue] (P4) to [out=225,in=45] (Qbar);

\node[measure,fill=green!10] (Q) at (-0.4,-0.4) {};
\node[] () at (-.9,-0.25) {$\mathcolor{red}{\Q[\theta]}$};
\draw[mid arrow,dashed,color=red] (Qbar) to [out=-90,in=60] (Q);

\end{tikzpicture}

\end{minipage}
\begin{minipage}{0.28\textwidth}
\centering
\begin{tikzpicture}[    mid arrow/.style={
        postaction={decorate,decoration={
            markings,
            mark=at position .8 with {\arrow{Latex[length=6.4pt, sep=-3.2pt]}}
    }}
  },]

\node[measure] (P1) at (0.5,-1.5) {};
\node[] () at (0.25,-1.5) {${\mathcolor{blue}{\P_1}}$};
\node[measure] (P2) at (-2,0) {};
\node[] () at (-2.35,0) {${\mathcolor{blue}{\P_2}}$};
\node[measure] (P3) at (-1.5,1.25) {};
\node[] () at (-1.85,1.25) {${\mathcolor{blue}{\P_3}}$};
\node[measure] (P4) at (1,0.75) {};
\node[] () at (0.65,1) {${\mathcolor{blue}{\P_4}}$};

\node[measure,fill=green!10] (Q) at (-0.4,-0.4) {};
\node[] () at (-.9,-0.25) {$\mathcolor{red}{\Q[\theta]}$};
\draw[mid arrow,dashed,color=blue] (P1) to [out=90,in=-25] (Q);
\draw[mid arrow,dashed,color=blue] (P2) to [out=-25,in=245] (Q);
\draw[mid arrow,dashed,color=blue] (P3) to [out=-25,in=90] (Q);
\draw[mid arrow,dashed,color=blue] (P4) to [out=225,in=45] (Q);

\end{tikzpicture}

\end{minipage}
\begin{minipage}{0.35\textwidth}
\centering
\begin{tikzpicture}[ mid arrow/.style={
        postaction={decorate,decoration={
            markings,
            mark=at position .8 with {\arrow{Latex[length=6.4pt, sep=-3.2pt]}}
    }}
  },]
\node[measure] (P1) at (0.5,-1.5) {};
\node[] () at (0.25,-1.5) {${\mathcolor{blue}{\P_1}}$};
\node[measure] (P2) at (-2,0) {};
\node[] () at (-2.35,0) {${\mathcolor{blue}{\P_2}}$};
\node[measure] (P3) at (-1.5,1.25) {};
\node[] () at (-1.85,1.25) {${\mathcolor{blue}{\P_3}}$};
\node[measure] (P4) at (1,0.75) {};
\node[] () at (0.65,1) {${\mathcolor{blue}{\P_4}}$};

\fill[green!10] (1,-2) -- (0.75, -1) -- plot[domain=0.5:2.5] (\x,{0.25*sin(0.5*\x*pi r)}) -- (2.5,-2) -- cycle;
\node[] () at (2,-1.5) {$\mathcal{C}$};

\node[measure,fill=green!10] (QC) at (0.9,-0.25) {};
\node[] () at (1.3,-0.45) {$\mathcolor{red}{\Q[\theta]}$};
\draw[mid arrow ,dashed,color=blue] (P1) to [out=0,in=-90] (QC);
\draw[mid arrow ,dashed,color=blue] (P2) to [out=-25,in=135] (QC);
\draw[mid arrow ,dashed,color=blue] (P3) to [out=-25,in=90] (QC);
\draw[mid arrow ,dashed,color=blue] (P4) to [out=0,in=45] (QC);

\end{tikzpicture}
\end{minipage}
\caption{A representation of problem \eqref{opt:P}, where $\C$ denotes the set of measures that attain the problem constraints. }
\label{fig:diagram-P}
\end{figure}

Note that if expert-k's model satisfies $\E^{\P^{(k)}}\big[\sup_{t\in[0,T]} |X_t|^2\big]<\infty$, then $\P^{(k)}\in \mcQ$. Similarly, if the average drift measure fulfils $\E^{\Q[\bmu]}\big[\sup_{t\in[0,T]} |X_t|^2\big]<\infty$, then $\Q[\bmu]\in \mcQ$.

\subsection{Value function}
Next, we rewrite the agent's optimisation problem \eqref{opt:P}, which is an optimisation problem over probability measures, as an optimisation problem over admissible drifts of $X$. For this we define the set of admissible drifts $\theta$ of $X$ as the set of processes which induce a probability measure $\Q[\theta]\in \mcQ$. That is the set of admissible drifts is given by
\[
\A := \left\{\left.\theta : \Omega \times [0,T]\to \R^d \;\right|\;  \Q[\theta]\in\mcQ \right\}.
\]
Thus, the set $\A$ is the set of processes that generate the measures in the set $\mcQ$. Moreover,  the set $\A$ and $\mcQ$ are isomorphic, one set is parametrised by $\F$-adapted processes, while the other is parametrised by probability measures. Furthermore, by Girsanov's Theorem, $W_t = \int_0^t \blambda_u\,\du + \bW_t$ is a $\Q[\theta]$-Brownian motion for each $k\in\mcK$, and therefore the $\Q[\theta]$-drift of $X$ is equal to $\theta$.%
        
To rewrite optimisation problem \eqref{opt:P} as an optimisation problem over the admissible strategies $\A$, we calculate the KL divergence from $\Q[\theta]$ to $\P^{(k)}$, and in particular the Radon-Nikodym derivative $\frac{\dQ[\theta]}{\dP^{(k)}}$. For any $\theta\in\A$, equivalently for any $\Q[\theta]\in\mcQ$, we have for all $k\in\mcK$, that
\begin{align*}
    \frac{\dQ[\theta]}{\dP^{(k)}}
    &= \frac{\dQ[\theta]}{\dQ[\bmu]}\;\frac{\dQ[\bmu]}{\dP^{(k)}}
    \\
    &= 
    e^{-\frac{1}{2}\int_0^T |\blambda_t|^2\dt-\int_0^T \blambda^\intercal_t\,\dbW_t}\;
    e^{-\frac12\int_0^T|\bgamma_t^{(k)}|^2\dt
    -\int_0^T\bgamma^{(k)\intercal} \dW^{(k)}_t}
    \\
    &= 
    e^{-\frac{1}{2}\int_0^T (|\blambda_t|^2+|\bgamma_t^{(k)}|^2)\dt
    -\int_0^T \blambda^\intercal_t\,(\dW_t^{(k)}+\bgamma^{(k)}_t\dt) -\int_0^T\bgamma^{(k)\intercal} \dW^{(k)}_t}    
    \\
    &= 
    e^{-\frac{1}{2}\int_0^T (|\blambda_t|^2+2\blambda^\intercal_t\,\bgamma^{(k)}_t+|\bgamma_t^{(k)}|^2)\dt
    -\int_0^T\big(\sigma_t^{-1}(\mu_t^{(k)}-\theta_t)\big)^\intercal\, \dW^{(k)}_t}
    \\
    &= 
    e^{-\frac{1}{2}\int_0^T (\mu_t^{(k)}-\theta_t)^\intercal\Sigma_t^{-1}(\mu_t^{(k)}-\theta_t)\,\dt
    -\int_0^T\big(\sigma_t^{-1}(\mu_t^{(k)}-\theta_t)\big)^\intercal\, \dW^{(k)}_t}.
\end{align*}%
Thus, letting $\lambda_t^{(k)}:=\sigma_t^{-1}(\mu_t^{(k)}-\theta_t)$, Girsanov's Theorem implies that $\left(W^{(k)}_t+\int_0^t \lambda_t^{(k)}\,dt\right)_{\tT}$ is a $\Q[\theta]$--Brownian motion.
Thus, for each $k\in\mcK$, the KL divergence becomes
\begin{align*}
\mcH\left[\Q[\theta]\,\big\| \,\P^{(k)}\right]
&=
\E^{\Q[\theta]}\left[\log\frac{\dQ[\theta]}{\dP^{(k)}}\right]
\\
&=
\E^{\Q[\theta]}\left[
-\tfrac{1}{2}\int_0^T (\mu_t^{(k)}-\theta_t)^\intercal\,
\Sigma_t^{-1}\,(\mu_t^{(k)}-\theta_t)
\dt-\int_0^T \big(\sigma_t^{-1}(\mu_t^{(k)}-\theta_t)\big)^\intercal\,\dW_t^{(k)}
\right]
\\
&=
\E^{\Q[\theta]}\left[
-\tfrac{1}{2}\int_0^T (\mu_t^{(k)}-\theta_t)^\intercal\,
\Sigma_t^{-1}\,(\mu_t^{(k)}-\theta_t)
\dt
\right.
\\
&\hspace*{5em} \left.
-\int_0^T \big(\sigma_t^{-1}(\mu_t^{(k)}-\theta_t)\big)^\intercal\,\left(\big(\dW_t^{(k)}+\lambda^{(k)}_t\,dt\big) - \lambda^{(k)}_t\,dt\right)
\right]
\\
&=
\E^{\Q[\theta]}\left[
\tfrac{1}{2}\int_0^T \Delta\theta^{(k)\intercal}_t\,\Sigma_t^{-1}\, \Delta\theta^{(k)}_t\;
\dt
\right],
\end{align*}
where, we introduce the notation $\Delta\theta^{(k)}(t,x):=\theta(t,x)-\mu^{(k)}(t,x)$, $\Delta\theta^{(k)}_t:=\Delta\theta^{(k)}(t,X_t)$, $\Sigma(t,x):=\sigma(t,x) \sigma(t,x)^\intercal$, and $\Sigma_t:=\Sigma(t,X_t)$.

Therefore, we rewrite optimisation problem \eqref{opt:P} in terms of an optimisation over $\theta$ as follows
\begin{equation}
\label{eqn:KL-in-terms-of-theta}
    \inf_{\theta\in\A}\sum_{k\in\mcK} \pi_k\,
    \E^{\Q[\theta]}\left[
    \tfrac12\int_0^T 
\Delta\theta^{(k)\,\intercal}_t\,\Sigma(t,X_t)^{-1}\,\Delta\theta^{(k)}_t\dt
    \right],
\end{equation}
subject to $\E^{\Q[\theta]}[\int_0^T g_u\,\du]=0$ and $\E^{\Q[\theta]}[f_T]=0$, where for simplicity of notation we write $g_u:=g(u, X_u)$ and $f_T:= f(X_T)$.

To solve this constrained optimisation problem, we introduce the associated Lagrangian with Lagrange multiplier $\eta:= (\eta_0, \eta_1)\in\R^2$, that is
\begin{align*}
    L_\eta[\theta]&(t,x) :=
\E^{\Q[\theta]}_{t,x}\left[\int_t^T
    \tfrac12\sum_{k\in\mcK} \pi_k \,
    \Delta\theta_u^{(k)\intercal}\,
    \Sigma_u^{-1} \Delta\theta_u^{(k)}
    \,
    \du
    + \eta_0\,\int_t^T g_u \,\du
    + \eta_1 \, f_T\, \right],
\end{align*}
where $\E_{t,x}^\Q[\cdot]$ denotes expectation conditional on $X_t=x$.
If we set $\eta=(0,0)$, then the problem reduces to finding the (pure) barycentre of the experts' models, i.e., without any constraints, which we discuss in \Cref{sec:barycentre}. That is, we first discuss the pure barycentre problem and then tackle the full optimisation problem \eqref{opt:P}.

For fixed $\eta\in\R^2$, we define the value function 
\begin{equation}\label{eq:optimal-performance}
L_\eta(t,x):=\inf_{\theta\in\A}L_\eta[\theta](t,x)\,.
\end{equation}
Note that for each $\eta$, the value function has an associated probability measure that attains the value function. Thus, when $\eta$ is chosen to bind the constraints the corresponding probability measure is the solution to optimisation problem \eqref{opt:P}.
Thus, in a first step, we find for fixed $\eta$ the value function $L_\eta(t,x)$ and in a second step find the optimal $\eta$ such that the constraints are fulfilled.

\begin{definition}
The following Mahalanobis--like distance is an important ingredient in the solution of optimisation problem \eqref{opt:P}
\begin{equation}
\label{eq:varsigma}
    \varsigma(t,x):= \tfrac12\sum_{k\in\mcK} \pi_k\,\dmu^{(k)}(t,x)^\intercal
    \,
    \Sigma^{-1}(t,x)
    \,
    \dmu^{(k)}(t,x),
\end{equation}
where $\dmu^{(k)}(t,x):=\mu^{(k)}(t,x)-\bmu(t,x)$.  
\end{definition}

\section{The pure barycentre}\label{sec:barycentre}

In this section, we address the pure barycentre problem, i.e., the case when the agent has no additional beliefs that they wish to incorporate. 
\begin{definition}[Barycentre measure]\label{def:barycentre}
 Consider optimisation problem \eqref{opt:P} without any constraints, whose solution we call the barycentre (probability) measure and denote by $\Q^\bary$, that is          
 \begin{equation*}
    \Q^\bary :=\argmin_{\Q\in\mcQ} \sum_{k\in \mcK}
    \pi_k\,\mcH\left[\Q\,\big\| \,\P^{(k)}\right]\,.
\end{equation*}
\end{definition}
This problem is equivalent to solving \eqref{eq:optimal-performance} with $\eta=(0,0)$. The theorem below provides one of our key results, i.e., the drift under the barycentre measure, given in \eqref{eqn:theta-0}, and the associated value function, given in \eqref{eqn:L-0}.
\begin{theorem}[Barycentre Drift and Value Function]
\label{thm:barycentre}
Let Assumptions \ref{asm:strong=sol-SDE} and \ref{asm:compatibility} be enforced. Define the function $L_0:[0,T]\times\R^d\to\R$, s.t.
\begin{equation}
\label{eqn:L-0}
    L_0(t,x) := -\log \E^{\Q[\bmu]}_{t,x}\left[ 
    \; e^{- \int_t^T \varsigma(u,X_u)\du} \; 
    \right], \qquad \forall (t,x) \in [0,T]\times\R^d\,.
\end{equation}
Suppose that  $L_0\in\C^{1,2}([0,T)\times\R^d;\R)\cap
\C^0([0,T]\times\R^d;\R)$ and has at most quadratic growth, i.e. there exists $C\in\R_+$ s.t. $|L_0(t,x)|\le C(1+|x|^2)$ for all $(t,x)\in[0,T]\times\R^d$. Next, define the process $\theta_0:=(\theta_{0,t})_{\tT}$  s.t.
\begin{equation}
\label{eqn:theta-0}
    \theta_{0,t} := \bmu_t-\Sigma_t\nabla_x L_0(t,X_t)\,,
\end{equation}
and suppose that
\begin{align}
\label{eqn:bary-assumption}
\E^{\Q[\bmu]}\left[e^{\frac12\int_0^T|\blambda_{0,u}|^2\du}\right]<+\infty\,, 
\quad \E^{\Q[\theta_0]}\left[\sup_{t\in[0,T]}|X_t|^2\right]<+\infty\,,
\end{align}
where $\blambda_{0,t}:=\sigma^{-1}_t\big(\bmu_t-\theta_{0,t}\big)$.
Then $\theta_0$ is the admissible control and $L_0$ is the value function associated with the pure barycentre problem --- problem \eqref{eq:optimal-performance} with $\eta=(0,0)$. Moreover, the probability measure that attains the minimum is $\Q^\bary=\Q[\theta_0]$.
\end{theorem}
\begin{proof}
First, note that $\theta_0$ is $\F$-adapted, and due to \eqref{eqn:bary-assumption}, we have that $\theta_0\in\A$ and that $\Q[\theta_0]$ is well-defined.

Next, we show that $L_0$ is the value function associated with the barycentre problem. To this end, define the function $\omega$, s.t. $\omega(t,x):=E^{\Q[\bmu]}_{t,x}\left[ 
    \; e^{- \int_t^T \varsigma(u,X_u)\du} \; 
    \right]$, i.e., $\omega=e^{-L_0}$.
Clearly, $\omega\in\C^{1,2}([0,T)\times\R^d;\R)\cap \C^0([0,T]\times\R^d;\R)$, which is inherited from $L_0$, and 
as $\Sigma$ is positive definite, it holds that 
\begin{equation}
    \E^{\Q[\bmu]}\left[\,e^{-\int_0^T \varsigma(u, X_u) \, \du}\,\right] \le 1\,.
    \label{eqn:varsigma}
\end{equation}
Therefore, the Feynman--Kac formula implies that $\omega$ satisfies the PDE
\begin{equation*}
\left\{
\begin{split}
    \partial_t \omega + 
        \bmu^\intercal\,\nabla_x \omega 
        + \tfrac12 {\Tr}\left(\Sigma\,\nabla_{x}^2 \omega\right) 
        -  \varsigma\,\omega
         &= 0,
        \\
        \omega(T,x) &= 1\,,
\end{split}
\right. 
\end{equation*}
where we suppress the $(t,x)$ arguments.
Hence, inserting $\omega=e^{-L_0}$, we have that $L_0$ satisfies the PDE
\begin{equation}
\left\{
\begin{split}
    \partial_t L_0 + 
        \bmu^\intercal\,\nabla_{x}L_0 
        + \tfrac12 {\Tr}\left(\Sigma\,\nabla_{x}^2 L_0\right)
        - \tfrac12 \nabla_{x}L_0^\intercal \Sigma \nabla_{x}L_0
        + \tfrac12 \sum_{k\in\mcK} \pi_k\,\dmu^{(k)^\intercal} \,\Sigma^{-1} \,\Delta \mu^{(k)}
         &= 0,
         \label{eqn:L_0-PDE-after-min-sub}
         \\
         L_0(T,x) &= 0\,,
\end{split}
\right.
\end{equation}
where we insert the representation for $\varsigma$ in \Cref{eq:varsigma}. We claim that \eqref{eqn:L_0-PDE-after-min-sub} is equivalent to
\begin{equation}
\left\{
\begin{split}
\label{eqn:HJB}
        \partial_t L_0 + \min_{\theta\in\R^d}
        \left\{\theta^\intercal\,\nabla_{x}L_0 
        + \tfrac12 {\Tr}\left(\Sigma\,\nabla_{x}^2 L_0\right) 
        + \tfrac12\sum_{k\in\mcK} \pi_k
        (\theta-\mu^{(k)})^\intercal\Sigma^{-1}(\theta-\mu^{(k)})
        \right\} &= 0,
        \\
        L_0(T,x) &= 0.
\end{split}
\right.
\end{equation}
This follows from noting that the minima above is attained by $\theta_0$ given in \eqref{eqn:theta-0}, and therefore
\allowdisplaybreaks
\begin{align}
    \min_{\theta\in\R^d} &
    \left\{\theta^\intercal\,\nabla_{x}L_0 
    + \tfrac12 {\Tr}\left(\Sigma\,\nabla_{x}^2 L_0\right) 
    + \tfrac12\sum_{k\in\mcK} \pi_k
    (\theta-\mu^{(k)})^\intercal\Sigma^{-1}(\theta-\mu^{(k)})
    \right\}
    \nonumber
    \\[1em]
    \begin{split}
        & =(\bmu -  \Sigma \,\nabla_{x}L_0)^\intercal\,\nabla_{x}L_0 
        + \tfrac12 {\Tr}\left(\Sigma\,\nabla_{x}^2 L_0\right) 
        \\
        & \quad + \tfrac12\sum_{k\in\mcK} \pi_k
        (\bmu -  \Sigma \,\nabla_{x}L_0 -\mu^{(k)})^\intercal\Sigma^{-1}(\bmu -  \Sigma \,\nabla_{x}L_0-\mu^{(k)}) 
    \end{split}
    \nonumber
    \\[1em]
    \begin{split}
        & = -\nabla_{x}L_0^\intercal\,\Sigma\,\nabla_{x}L_0 
        + \tfrac12 {\Tr}\left(\Sigma\,\nabla_{x}^2 L_0\right) 
        +\bmu^\intercal \nabla_{x}L_0 
        \\
        & \quad
        + \tfrac12\sum_{k\in\mcK} \pi_k
        \left\{
        \dmu^{(k)\intercal}\Sigma^{-1}\dmu^{(k)}   
        +\nabla_{x}L_0^\intercal\,\Sigma \,\nabla_{x}L_0
        + 2\dmu^{(k)\intercal} \nabla_{x}L_0
        \right\}\,.
    \end{split}
    \nonumber
    \\[1em]
    &
        = -\tfrac12\nabla_{x}L_0^\intercal\,\Sigma\,\nabla_{x}L_0 
        + \tfrac12 {\Tr}\left(\Sigma\,\nabla_{x}^2 L_0\right) 
        +\bmu^\intercal \nabla_{x}L_0 
                + \tfrac12\sum_{k\in\mcK} \pi_k
        \dmu^{(k)^\intercal}\Sigma^{-1}\dmu^{(k)}  \,.%
    \label{eqn:min-term}
\end{align}%

We next argue that the value associated with an arbitrary control is lower bounded by $L_0(t,x)$.
For this purpose,  take an arbitrary $\theta\in\A$ and $s\in[t,T)$, and a stopping time $\tau_n=\inf\{u\ge t : |X_u|>n\}$, for $n\in\Z_+$, $n<\infty$. From Dynkin's formula, we have that
\begin{equation}
    \E^{\Q[\theta]}_{t,x}\Big[ L_0(s\wedge\tau_n, X_{s\wedge\tau_n})\Big]
    = L_0(t,x) + \E^{\Q[\theta]}_{t,x}\left[ \int_t^{s\wedge\tau_n} \left\{\partial_t L_0(u, X_u) + \L^\theta L_0(u,X_u)\right\}du\right],
    \label{eqn:dynkin}
\end{equation}
where the operator $\L^\theta$ is the $\Q[\theta]$-infinitesimal generator, and acts on functions $h:\R_+\times\R^d\to\R$ as follows $\L^\theta h:=\theta^\intercal\,\nabla_{x}h 
        + \tfrac12 {\Tr}\left(\Sigma\,\nabla_{x}^2 h\right)$.
Next, due to \eqref{eqn:HJB} and  \eqref{eqn:min-term}, and as $\theta$ is arbitrary, we have that
\[
\partial_t L_0(t,X_t) + \L^\theta L_0(t,X_t) + \tfrac12\sum_{k\in\mcK} \pi_k \;\Delta\theta^{(k)\intercal}_t\;\Sigma^{-1}_t\;\Delta\theta^{(k)}_t
        \ge 0,
\]
Combined with \eqref{eqn:dynkin}, we therefore have that
\begin{equation}  
\label{eqn:for-dominated-conv}
\begin{split}
    & \E^{\Q[\theta]}_{t,x}\Big[ L_0(s\wedge\tau_n, X_{s\wedge\tau_n})\Big]
    \\
    & \hspace{2em} \ge L_0(t,x) - \E^{\Q[\theta]}_{t,x}\left[ \int_t^{s\wedge\tau_n}
    \tfrac12\sum_{k\in\mcK} \pi_k \; \Delta\theta^{(k)\intercal}_u\;\Sigma^{-1}_u\;\Delta\theta^{(k)}_u
    du\right],    
\end{split}
\end{equation}
We wish to take the limit as $n\to\infty$ and interchange the limit and the expectation. For this purpose, we argue that each side of the above expression is $\Q[\theta]$-integrable. First, for the term under the conditional expectation on the rhs, we have
\begin{multline}
    \E^{\Q[\theta]}\left[ \left| \int_t^{s\wedge\tau_n}
    \tfrac12\sum_{k\in\mcK} \pi_k \; \Delta\theta^{(k)\intercal}_u\;\Sigma^{-1}_u\;\Delta\theta^{(k)}_u
    du\right| \right]
    \\
    \le 
    \E^{\Q[\theta]}\left[ \left| \int_0^{T}
    \tfrac12\sum_{k\in\mcK} \pi_k \; \Delta\theta^{(k)\intercal}_u\;\Sigma^{-1}_u\;\Delta\theta^{(k)}_u
    du\right| \right] < \infty,
\end{multline}
where the first inequality follows as $\Sigma$, and hence $\Sigma^{-1}$, are strictly positive definite and where the last inequality follows as $\theta\in\A$ and, therefore, $\Q[\theta]\in\mcQ$.
Next, due to the quadratic growth condition of $L_0$, we have that
\[
\E^{\Q[\theta]}\left[\Big|L_0(s\wedge\tau_n, X_{s\wedge\tau_n})\Big|\right] \le C \left(1 + \E^{\Q[\theta]}\left[\sup_{u\in [t,T]} |X_u|^2\right]\right)
< \infty,
\]
as $\Q[\theta]\in\mcQ$. Consequently, we may apply the dominated convergence theorem to take the limit as $n\to\infty$ and interchange the limit and expectation in \eqref{eqn:for-dominated-conv} to obtain
\begin{equation}
\begin{split}
        & \E^{\Q[\theta]}_{t,x}\Big[ L_0(s, X_{s})\Big]
    \\
    & \hspace{2em} \ge L_0(t,x) - \E^{\Q[\theta]}_{t,x}\left[ \int_t^{s}
    \tfrac12\sum_{k\in\mcK} \pi_k \; \Delta\theta^{(k)\intercal}_u\;\Sigma^{-1}_u\;\Delta\theta^{(k)}_u
    du\right]    
\end{split}
\end{equation}
for all $s\in[t,T)$.

By continuity of $L_0$, taking the limit as $s\uparrow T$, we have
\begin{subequations}
\label{eqn:in-eq}
\begin{align}
   0&= \E^{\Q[\theta]}_{t,x}\Big[ L_0(T, X_{T})\Big]
    \\
    &  \ge L_0(t,x) - \E^{\Q[\theta]}_{t,x}\left[ \int_t^{T}
    \tfrac12\sum_{k\in\mcK} \pi_k \; \Delta\theta^{(k)\intercal}_u\;\Sigma^{-1}_u\;\Delta\theta^{(k)}_u
    du\right]    
\end{align}%
\end{subequations}%
Therefore,
\[
L_0(t,x) \le 
\E^{\Q[\theta]}_{t,x}\left[ \int_t^{T}
    \tfrac12\sum_{k\in\mcK} \pi_k \; \Delta\theta^{(k)\intercal}_u\;\Sigma^{-1}_u\;\Delta\theta^{(k)}_u
    du\right]
= L_0[\theta](t,x).
\]
So that $L_0(t,x)$ is a lower bound for the value associated with any arbitrary $\theta\in\A$. 

Finally, we show that $L_0(t,x)\ge L_0[\theta_0](t,x)$. To this end, using a similar localization argument as above, but starting with the specific control $\theta_0$, we have, by construction of $\theta_0$, that
\[
\partial_t L_0(t, X_t) + \L^{\theta_0} L_0(t,X_t) + \tfrac12\sum_{k\in\mcK} \pi_k \;\Delta\theta^{(k)\intercal}_t\;\Sigma^{-1}_t\;\Delta\theta^{(k)}_t = 0,
\]
and that, similar to \eqref{eqn:in-eq} but now with equality due to the above, we have
\[
0 = L_0(t,x) - \E^{\Q[\theta_0]}_{t,x}\left[ \int_t^{T}
    \tfrac12\sum_{k\in\mcK} \pi_k \; \Delta\theta^{(k)\intercal}_u\;\Sigma^{-1}_u\;\Delta\theta^{(k)}_u
    du\right].
\]
Therefore, re-arranging, we obtain $L_0(t,x)=L_0[\theta_0](t,x)$. Combined with the earlier inequality, $L_0(t,x)\le L_0[\theta](t,x)$ for all $\theta\in\A$, we conclude the proof.
\end{proof}

The RN derivative of the barycentre model to the average drift model admits a succinct representation as presented next.
\begin{proposition}[Pure barycentre Measure Change]
\label{prop:bary}
Under the assumptions stated in \Cref{thm:barycentre}, the barycentre measure exists, is unique, and its RN derivative has representation
\begin{equation}
\label{eqn:bary-rep}
    \frac{\d \Q^\bary}{\dQ[\bmu]} 
    = 
    \frac{e^{- \int_0^T \varsigma(t,X_t) \,\dt }}{\E^{\Q[\bmu]}\left[
    e^{- \int_0^T \varsigma(t,X_t)\,\dt}\right]}\;.
\end{equation}
\end{proposition}
\begin{proof}
Existence and uniqueness follows as the KL divergence is strictly convex and coercive\footnote{The KL being coercive can be seen as following from Pinsker's inequality which states $KL(\Q \| \P)\ge 2 \|\Q-\P\|_{TV}$ where $\|\cdot\|_{TV}$ is the total variation distance. Thus, with $\P$ fixed, as $ \|\Q-\P\|_{TV}\to\infty$ so does $KL(\Q \| \P)$. }.
From \Cref{thm:barycentre} we have that the Lagrangian $L_0(t,x) = -\log \E^{\Q[\bmu]}_{t,x}\left[ 
    \; e^{- \int_t^T \varsigma(u,X_u)\, \du} \; 
    \right]$ and
by  \Cref{eqn:varsigma}, $L_0(t,x)$ is finite for all $t$ and $x$.
The representation \eqref{eqn:bary-rep} follows along the same lines as in Theorem 2.7 in \cite{Jaimungal2024SICON}, however, for completeness, we provide a self-contained proof below.

Define the process $(\tomega_t)_{\tT}$, s.t. $\tomega_t:= E^{\Q[\bmu]}_{t,X_t}[ e^{-\int_0^T \varsigma(u,X_u)\,\du}]$. We may write $\tomega_t=e^{-\int_0^t \varsigma(u,X_u)\,\du}\omega(t,X_t)$, where $\omega(t,x):=E^{\Q[\bmu]}_{t,x}[ e^{-\int_t^T \varsigma(u,X_u)\,\du}]$. By the assumptions in \Cref{thm:barycentre}, we have that i) $\tomega$ is a $\Q[\bmu]$-martingale and ii)  $\omega\in\C^{1,2}([0,T)\times\R^d;\R)$. Therefore, applying Ito's lemma, we have 
\begin{multline}
    \d\tomega_t = e^{-\int_0^t \varsigma(u,X_u)\,du}\Big\{
    \Big[
    \partial_t\omega(t,X_t)+\bmu^\intercal \nabla_x\omega(t,X_t)
    \\
    + \tfrac12 \Tr(\Sigma(t,X_t)\nabla_x\omega(t,X_t)
    \Big]\,\dt
    +\nabla_x\omega(t,X_t)^\intercal\sigma\,\dbW_t \Big\} 
    \\
    -e^{-\int_0^t \varsigma(u,X_u)\,\du}\,\varsigma(t,X_t)\, \omega(t,X_t)\,\dt\,.
\end{multline}
As $\tomega$ is a  $\Q[\bmu]$-martingale, we must have
\begin{equation}
\label{eqn:tomega-martingale}
    \partial_t\omega(t,X_t)+\bmu^\intercal \nabla_x\omega(t,X_t) 
    \\
    + \tfrac12 \Tr(\Sigma(t,X_t)\nabla_x\omega(t,X_t)
    =\varsigma(t,X_t)\, \omega(t,X_t)\,.
\end{equation}

Next, introduce the process $V:=1/\tomega$, then by Ito's lemma, it follows that (where we write $\varsigma_t=\varsigma(t,X_t)$ and suppress the arguments $(t,X_t)$)
\begin{equation}
\begin{split}
\d V_t  &= -\frac{e^{\int_0^t\varsigma_u\,\du}}{\omega^2}
    \Big( \partial_t\omega+\bmu^\intercal \nabla_x\omega
    + \tfrac12 \Tr\big(\Sigma\nabla_x\omega\big)
\Big)\dt 
- \frac{e^{\int_0^t\varsigma_u\,\du}}{\omega^2} (\nabla_x\omega)^\intercal \,\sigma\,\dbW_t
\\
&\quad 
+     \frac{e^{\int_0^t\varsigma_u\,\du}}{\omega^3} (\nabla_x\omega)^\intercal\,\Sigma\,\nabla_x\omega\,\dt 
+ \varsigma\,\frac{e^{\int_0^t\varsigma_u\,\du}}{\omega}\,\dt\,.
\end{split}
\end{equation}
Using the identity \eqref{eqn:tomega-martingale}, we obtain
\begin{align}
\d V_t &= -\frac{e^{\int_0^t\varsigma_u\,\du}}{\omega}\varsigma \,\dt- \frac{e^{\int_0^t\varsigma_u\,\du}}{\omega^2} (\nabla_x\omega)^\intercal \,\sigma\,\dbW_t 
\\
&\quad 
+     \frac{e^{\int_0^t\varsigma_u\,\du}}{\omega^3} (\nabla_x\omega)^\intercal\,\Sigma\,\nabla_x\omega\,\dt 
+ \varsigma\,\frac{e^{\int_0^t\varsigma_u\,\du}}{\omega}\,\dt
\\
&= - \frac{e^{\int_0^t\varsigma_u\,\du}}{\omega^2} (\nabla_x\omega)^\intercal \,\sigma\,\dbW_t 
+     \frac{e^{\int_0^t\varsigma_u\,\du}}{\omega^3} (\nabla_x\omega)^\intercal\,\Sigma\,\nabla_x\omega\,\dt \,.
\end{align}

Introduce the process $\phi= Z\,V$, where 
\[
Z_t:=e^{-\frac12\int_0^t |\lambda_u|^2\,\dt - \int_0^t\lambda_u^\intercal\,\dbW_u}\,,
\quad \text{ and } \quad
\lambda_t:=\sigma(t,X_t)^\intercal\frac{\nabla_x\omega(t,X_t)}{\omega(t,X_t)},
\]
and where $\bW$ is a $\Q[\bmu]$-Brownian motion. We next establish that $\phi$ is a constant.

By the assumptions in \Cref{thm:barycentre}, $Z$ is a $\Q[\bmu]$-martingale, and from Ito's lemma $dZ_t=Z_t\,\lambda_t^\intercal\,\d\bW_t$. From here, it follows that 
\begin{align}
    \d \phi_t &= \d Z_t \,V_t + Z_t\,\d V_t + \d[Z,V]_t
    \\
    \begin{split}
        &= V_t\,Z_t\,\lambda_t^\intercal\,\dbW_t -Z_t \frac{e^{\int_0^t\varsigma_u\,\du}}{\omega^2} (\nabla_x\omega)^\intercal \,\sigma\,\dbW_t 
        \\
        &\qquad 
        +Z_t\, \frac{e^{\int_0^t\varsigma_u\,\du}}{\omega^3} (\nabla_x\omega)^\intercal\,\Sigma\,\nabla_x\omega\,\dt
     - Z_t\,\frac{e^{\int_0^t\varsigma_u\,\du}}{\omega^2} (\nabla_x\omega)^\intercal \,\sigma \lambda_t \,\dt\,.
    \end{split}
    \\
    \begin{split}
        &= 
        \frac{e^{\int_0^t\varsigma_u\,\du}}{\omega}\,Z_t\,\frac{\nabla_x\omega^\intercal}{\omega}\sigma \,\dbW_t -Z_t \frac{e^{\int_0^t\varsigma_u\,\du}}{\omega^2} (\nabla_x\omega)^\intercal \,\sigma\,\dbW_t 
        \\
        &\qquad 
        +Z_t\,\frac{e^{\int_0^t\varsigma_u\,\du}}{\omega^3} (\nabla_x\omega)^\intercal\,\Sigma\,\nabla_x\omega
        \,\dt
         - Z_t\,\frac{e^{\int_0^t\varsigma_u\,\du}}{\omega^3} (\nabla_x\omega)^\intercal \,\Sigma \nabla_x\omega \,\dt
    \end{split}
    \\
    &=0\,.
\end{align}
Consequently, $\phi_t=C$ for some $C\in\R$. Next, as $C=\phi_T=Z_T\,V_T= Z_T\,e^{\int_0^T\varsigma_u\,\du}$,  we have that
\begin{equation}
    \frac{\dQ^\bary}{\dQ[\bmu]} = e^{-\frac12\int_0^T |\lambda_u|^2\,\dt - \int_0^T\lambda_u^\intercal\,\dbW_u} =Z_T=C\,e^{-\int_0^T\varsigma_u\,\du}\,.
\end{equation}
Moreover, by assumption, $\E^{\Q[\bmu]}[e^{\frac12\int_0^T |\lambda_u|^2\,\dt}]<\infty$, and hence from Theorem 1 in \cite{novikov1973identity}, we have that $\E^{\Q[\bmu]}[e^{-\frac12\int_0^T |\lambda_u|^2\,\dt - \int_0^T\lambda_u^\intercal\,\dbW_u}]=1$.  Continuing, by assumption, $\E^{\Q[\bmu]}[e^{-\int_0^T\varsigma_u\,\du}]<\infty$, we deduce that $C=1/\E^{\Q[\bmu]}[e^{-\int_0^T\varsigma_u\,\du}]$, and the result follows.
\end{proof}

We remark that the RN derivative of the barycentre to the average drift model is of exponential type, but different to the solution to the classical entropy maximisation, see e.g., \cite{Csiszar1975AP}. The exponent is a time integral over the weighted deviation between each expert's drift to the average drift $\bmu(t,x)$ scaled by the instantaneous covariance matrix $\Sigma(t,x)$ --- akin to a Mahalanobis distance between models. Moreover, note that $\Q[\bmu] = \Q^\bary$ if $i)$ all models are the same, or $ii)$ if $\Sigma(t,x)$ and $\mu^{(k)}(t,x)$, for all $k \in \mcK$, are functions of time only.

\subsection{The case of two experts}\label{sub-sec:bary-two}

Consider a scenario where there are two experts and the agent uses $\pi^{(1)}=1-\ep$ and $\pi^{(2)}=\ep$, the following result shows how the barycentre drift moves from $\mu^{(1)}$ towards $\mu^{(2)}$ for small values of $\ep$. That is, how expert model--1 is perturbed in the direction of expert model--2 through the barycentre.

\begin{proposition}[Asymptotic Expansion.]
\label{prop:asymptotic-expansion-L}
Suppose the assumptions in \Cref{thm:barycentre} hold. Consider the case of two expert models and set $\pi_1=(1-\ep)$ and $\pi_2=\ep$ for $\ep\in[0,1]$. Further, assume that
\begin{equation}
\E^{\Q[\bmu]}\left[\int_0^T \!\!\! \left(\left|\Gamma_u\right|+ \left|\Gamma_u\,\Upsilon_u\right|\right)
    \,\du    \right] < \infty,
\label{eqn:Gamma-Upsilon-assumption}
\end{equation}
where $(\Gamma_t)_{\tT}$ is the process s.t. $\Gamma_t:=\frac12\dmu_t\,\Sigma^{-1}_t\,\dmu_t$ with $\dmu_t=\mu^{(2)}_t-\mu^{(1)}_t$ and $(\Upsilon_t)_{\tT}$ is the process s.t. $\Upsilon_t:=\E^{\Q[\bmu]}_{t}\left[  \int_t^T \;\Gamma_u\,\du \; \right]$. Then, the value function admits the asymptotic expansion
\begin{equation}
L_0(t,x) = \ep\, \E^{\Q[\bmu]}_{t,x}\left[  \int_t^T \;\Gamma_u\,\du\; \right]+o(\ep) \,.
\label{eqn:L0-ep}
\end{equation}
\end{proposition}
\begin{proof}
To prove this, we show that 
\[
\omega(t,x):=e^{-L_0(t,x)} =  
\E^{\Q[\bmu]}_{t,x}\left[ e^{-\int_t^T \varsigma_u\,\du}\right]
=
1-
\ep\, \E^{\Q[\bmu]}_{t,x}\left[  \int_t^T \Gamma_u\,\du\; \right]+o(\ep) \,,
\]
where the first equality is the definition of $\omega(t,x)$ and the second follows from \eqref{eqn:L-0} in \Cref{thm:barycentre}, thus we proceed to show the last equality.
Once established,  \eqref{eqn:L0-ep} follows immediately from elementary results.

To this end, define 
\[
\tomega(t,x):=1- \ep\, \E^{\Q[\bmu]}_{t,x}\left[  \int_t^T \;\Gamma_u\,\du\; \right]
\]
and $r(t,x):=\omega(t,x)-\tomega(t,x)$. Then, due to the assumptions, from Feynman-Kac we have that $\omega$ and $\tomega$ satisfy the PDEs
\begin{subequations}
\begin{align}
\left\{
\begin{aligned}
    \partial_t \omega + \bmu^\intercal\nabla\omega + \tfrac12 \Tr\left(\Sigma\,\nabla_{x}^2 \omega\right) - \varsigma\, \omega&= 0,
    \\[0.5em]
    \omega(T,x) &= 1,        
\end{aligned}
\label{eqn:omege-PDE}
\right.
\\[1em]
\left\{
\begin{aligned}
    \partial_t \tomega + \bmu^\intercal\nabla\tomega + \tfrac12 \Tr\left(\Sigma\,\nabla_{x}^2 \tomega\right) - \ep\, \Gamma &= 0,
    \\[0.5em]
     \tomega(T,x) &= 1,
\end{aligned}
\label{eqn:tomega-PDE}
\right.
\end{align}%
\end{subequations}%
where $\Gamma(t,x):=\frac12\dmu(t,x)\,\allowbreak\Sigma^{-1}(t,x)\,\dmu(t,x)$ and, with a slight abuse of notation when we omit the subscript on $\Gamma$,   --- we further omit the arguments $(t,x)$ the above PDEs.

Note that, for the case of two expert models and the choice of $\pi$ in the proposition statement, we have
\[
\varsigma = \tfrac12 \left((1-\ep)(\mu^{(1)} - \bmu)\Sigma^{-1}(\mu^{(1)}-\bmu)
+\ep (\mu^{(2)} - \bmu)\Sigma^{-1}(\mu^{(2)}-\bmu)\right)
\]
and $\bmu=(1-\ep)\mu^{(1)} + \ep\,\mu^{(2)} = \mu^{(1)}+\ep\dmu$, where $\dmu=\mu^{(2)}-\mu^{(1)}$.
Hence,
\[
\mu^{(1)}-\bmu = -\ep\dmu,
\qquad
\mu^{(2)}-\bmu = \mu^{(2)}-\mu^{(1)}-\ep\dmu
=(1-\ep)\dmu\,.
\]
Plugging these representation of $\mu^{(1)}$ and $\mu^{(2)}$ into 
$\varsigma$, we obtain
\begin{equation}
    \varsigma =\tfrac12\left( (1-\ep)\ep^2\dmu\,\Sigma^{-1}\,\dmu
    +\ep (1-\ep)^2 \dmu\,\Sigma^{-1}\,\dmu \right)
    = \ep(1-\ep)\;\Gamma\,.
\end{equation}    
 
Inserting $\omega = \tomega+r$ into \eqref{eqn:omege-PDE}, and using \eqref{eqn:tomega-PDE}, we find that
\begin{align}
\begin{split}
        0&=\partial_t \left(\tomega+r\right) + \bmu^\intercal\nabla\left(\tomega+r\right) + \tfrac12 \Tr\left(\Sigma\nabla_{x}^2 \left(\tomega+r\right)\right) 
        - \ep(1-\ep) \Gamma \left(\tomega+r\right)
        \\
        &=
        \partial_t r + \bmu^\intercal\nabla_{x}r + \tfrac12 \Tr\left(\Sigma\nabla_{x}^2 r\right)
        - \ep(1-\ep) \Gamma\,r
        +\ep \big( \Gamma(1- \tomega)
        + \ep\Gamma\tomega \big)       
        \label{eqn:r-PDE-I}
\end{split}
\end{align}
Next, define $\Upsilon(t,x):=\E^{\Q[\bmu]}_{t,x}\left[  \int_t^T \;\Gamma_u\,\du \; \right]$ and note that $\tomega(t,x)=1-\ep\,\Upsilon(t,x)$, hence, the PDE \eqref{eqn:r-PDE-I} may be written as
\begin{equation}
\begin{split}
\partial_t r + \bmu^\intercal\nabla_{x}r + \tfrac12 \Tr\left(\Sigma\,\nabla_{x}^2 r\right)
- \ep(1-\ep) \Gamma\,r
+\ep^2\big(\Gamma+ (1-\ep)\,\Gamma\,\Upsilon\big)  = 0\,,        
\end{split}
\end{equation}
with terminal condition $r(T,x) = 0$. Feynman-Kac then gives
\begin{align}
    r(t,x) =
    \ep^2\;
    \E^{\Q[\bmu]}_{t,x}\left[
    \int_t^T e^{-\ep(1-\ep)\int_u^T \Gamma_s\,\ds}
    \big(\Gamma_u+ (1-\ep)\,\Gamma_u\,\Upsilon_u\big)
    \,\du\right]\,.
\end{align}

Recall that $\Gamma_t=\frac12\dmu_t\Sigma^{-1}\dmu_t$, and as $\Sigma$ is strictly positive definite, $\Gamma_t>0$ a.s. Hence, for $\ep\in[0,1]$, we have that 
\begin{align}
&\hspace*{-3em}
\left|
\E^{\Q[\bmu]}_{t,x}\left[
    \int_t^T e^{-\ep(1-\ep)\int_u^T \Gamma_s\,\ds}
    \big(\Gamma_u+ (1-\ep)\,\Gamma_u\,\Upsilon_u\big)
    \,\du\right]
\right|
\\
&\le
\E^{\Q[\bmu]}_{t,x}\left[
\left|
    \int_t^T e^{-\ep(1-\ep)\int_u^T \Gamma_s\,\ds}
    \big(\Gamma_u+ (1-\ep)\,\Gamma_u\,\Upsilon_u\big)
    \,\du
    \right|\right]
\\
&\le 
\E^{\Q[\bmu]}_{t,x}\left[
     \int_0^T \big(|\Gamma_u|+ \left|\Gamma_u\,\Upsilon_u\right|\big)
    \,\du
    \right].
\end{align}
Therefore, due to the bound \eqref{eqn:Gamma-Upsilon-assumption},  we have that
\begin{align}
\lim_{\ep\downarrow0}
\left|\frac{r(t,x)}{\ep}\right|
&=
\lim_{\ep\downarrow0}
\ep
\left|
\E^{\Q[\bmu]}_{t,x}\left[
    \int_t^T e^{-\ep(1-\ep)\int_u^T \Gamma_s\,\ds}
    \big(\Gamma_u+ (1-\ep)\,\Gamma_u\,\Upsilon_u\big)
    \,\du\right]
\right|
= 0\,.
\end{align}
and the result follows.~
\end{proof}

\begin{corollary}[Perturbed Drift.]
Let the assumptions in \Cref{prop:asymptotic-expansion-L} hold. Then the drift on the barycentre model with two experts and weights $\pi^{(1)} =1-\ep $, $\pi^{(2)} = \ep$ is
\begin{equation}
    \theta_{0,t} = \mu^{(1)}_t + \ep\left( 
    \Delta\mu_t  -  \Sigma_t \;\nabla_x\left( \E_{t,x}^{\Q[\bmu]}\left[\int_t^T \Delta\mu_s^\intercal\, \Sigma_s^{-1} \,\Delta\mu_s\,\ds\right] \right)\right)
    +o(\ep).
\end{equation}
    
\end{corollary}
\begin{proof}
    The result follows immediately from the optimal drift \eqref{eqn:theta-0} from \Cref{thm:barycentre} and \Cref{prop:asymptotic-expansion-L}.
\end{proof}

\subsection{Ornstein-Uhlenbeck experts' processes}\label{sub-sec:bary-OU}

In this section, we provide an example for a class of Ornstein-Uhlenbeck (OU) processes that allow for semi-explicit solutions --- up to the solution of a Ricatti ordinary differential equation (ODE). To this end, 
Suppose that $d=1$ and that $\mu^{(k)}(t,x) = a^{(k)}_t-b^{(k)}_t x$ and $\sigma(t,x)=\sigma_t\ge\ep$ (for some $\ep>0$) where $a^{(k)}_t$, $b^{(k)}_t$, and $\sigma_t$ are bounded  deterministic functions of time, for all $k\in\mcK$. Then we have that
\begin{equation}
\varsigma(t,x) = \frac1{2\sigma^2} \sum_{k\in\mcK}  \pi_k \big(\alpha_t^{(k)} - \beta^{(k)}_t x  \big)^2\,,
\end{equation}
where $\alpha_t^{(k)} :=a^{(k)}_t-\oa_t$, $\oa_t:=\sum_{k\in\mcK}a^{(k)}_t$, 
$\beta_t^{(k)} :=b^{(k)}_t-\ob_t$, and $\ob_t:=\sum_{k\in\mcK}b^{(k)}_t$. We further assume, without loss of generality, that at least one $\beta^{(k)} \ne 0$ --- otherwise, $\varsigma$ is a constant and, hence, from \Cref{prop:bary}, $\Q^\bary = \Q[\bmu]$, i.e., the barycentre model is the one with the drift equal to the average of the experts' drifts.

We next aim to find $L_0(t,x)$ in \eqref{eqn:L-0} and subsequently, the optimal drift $\theta_0$ in \eqref{eqn:theta-0}. 
\begin{proposition}
\label{prop:OU}
    The drift in the barycentre model for the expert models specified above is given by
    \begin{equation}
        \theta_{0,t} = \bmu_t + \sigma_t^2 (\mfb_t + \mfc_t\,X_t),
    \end{equation}
    where $\mfc_t$ solves the Ricatti equation 
    \begin{align}
        \dot\mfc_t- \ob_t\,\mfc_t+\tfrac12\sigma^2_t\mfc_t^2-\zeta_t
        &=0, & \mfc_T=0
    \end{align}
    $\zeta_t:=\frac{1}{2\sigma^2}\displaystyle\sum_{k\in\mcK}\pi_k\,\big(\beta_t^{(k)}\big)^2$ and $\mfb_t$ is given by
    \begin{equation}
        \mfb_t  
       = 
       \int_t^T \left(\oa_s\,\mfc_s  
        +\frac{1}{\sigma^2_s}
        \sum_{k\in\mcK}\pi_k \,\alpha_s^{(k)}\,\beta_s^{(k)}\right)e^{\int_t^s (\sigma^2_u\mfc_u-\ob_u)\,\du}\,
        \ds.
    \end{equation}
    Moreover, $\mfc_t$ and $\mfb_t$ remain bounded on the domain $t\in[0,T]$.
\end{proposition}
\begin{proof}
Define 
$\omega(t,x):=\E^{\Q[\bmu]}_{t,x}\Big[e^{-\int_t^T \varsigma(t,X_t)\,dt}\Big]$, Feynman-Kac formula implies that $\omega(t,x)$ satisfies the PDE (where we suppress the arguments $(t,x)$)
\begin{equation}
    \partial_t \omega + (\oa_t-\ob_t\,x)\partial_x \omega + \tfrac12 \sigma^2 \partial_{xx} \omega = \varsigma\,\omega, \qquad \omega(T,x)=1\,.
    \label{eqn:OU-PDE}
\end{equation}
As this is an affine PDE, it admits the solution of the form $\omega(t,x)=e^{\mfa_t+\mfb_t\,x+\frac12\mfc_t\,x^2}$, where $\mfa_t$, $\mfb_t$, and $\mfc_t$ are deterministic functions of time, such that $\mfa_T=\mfb_T=\mfc_T=0$.
Plugging this form into \eqref{eqn:OU-PDE}, we have
\begin{equation}
    \begin{split}
    (\dot\mfa_t + \dot\mfb_t\,x + \dot\mfc_t\,x^2) &+ 
    (\oa_t-\ob_t\,x)\,(\mfb_t+\mfc_t\,x)
    \\
    &+ \tfrac12\sigma^2_t\, \big( \mfc_t + (\mfb_t+\mfc_t\,x)^2\big) = \frac{1}{2\sigma^2_t}\sum_{k\in\mcK}\pi_k\Big(\alpha_t^{(k)}-\beta_t^{(k)}x\Big)^2.            
    \end{split}
\end{equation}
Collecting terms that are constant in $x$, terms that are linear in $x$, and terms that are quadratic in $x$, we have
\begin{align}
\begin{split}
    &  \left(\dot\mfc_t- \ob_t\,\mfc_t+\tfrac12\sigma^2_t\mfc_t^2-\frac{1}{2\sigma^2_t}\sum_{k\in\mcK}\pi_k\,\big(\beta_t^{(k)}\big)^2\right)\,x^2
    \\
    & \qquad +
    \left(\dot\mfb_t-\ob_t\,\mfb_t+\oa_t\,\mfc_t + \sigma^2_t \mfb_t\,\mfc_t
    +\frac{1}{\sigma^2_t}
    \sum_{k\in\mcK}\pi_k \,\alpha_t^{(k)}\,\beta_t^{(k)}\right)\,x
    \\
    & \qquad\qquad
    +    \left(\dot\mfa_t + \oa_t\,\mfb_t + \tfrac12\sigma^2_t(\mfc_t+\mfb_t^2) - \frac{1}{2\sigma^2_t}\sum_{k\in\mcK}\pi_k\,\big(\alpha^{(k)}_t\big)^2\right) = 0\,.    
\end{split}
\end{align}
As this must hold for all $t,x$, we obtain the system of ordinary differential equations
\begin{align}
\dot\mfc_t- \ob_t\,\mfc_t+\tfrac12\sigma^2_t\mfc_t^2-\frac{1}{2\sigma^2_t}\sum_{k\in\mcK}\pi_k\,\big(\beta_t^{(k)}\big)^2
&=0, & \mfc_T=0
\label{eqn:c-ode}
\\
\dot\mfb_t+(\sigma^2_t\,\mfc_t-\ob_t)\,\mfb_t+\oa_t\,\mfc_t  
    +\frac{1}{\sigma^2_t}
    \sum_{k\in\mcK}\pi_k \,\alpha_t^{(k)}\,\beta_t^{(k)}
&=0,
& \mfb_T=0
\\
\dot\mfa_t + \oa_t\,\mfb_t + \tfrac12\sigma^2_t(\mfc_t+\mfb_t^2) - \frac{1}{2\sigma^2_t}\sum_{k\in\mcK}\pi_k\,\big(\alpha^{(k)}_t\big)^2
&=0
& \mfa_T=0.
\end{align}
We next prove that the $\mfc_t$, which satisfies the Ricatti ODE \eqref{eqn:c-ode}, remains finite on the domain $\tT$, and subsequently so do $\mfb_t$ and $\mfa_t$.

Accordingly, define $\alpha_t := \int_t^T \ob_s\,\ds$, then
\[
\tfrac{\d}{\dt}(e^{\alpha_t} \mfc_t)
= e^{\alpha_t}\left(-\ob_t\,\mfc_t
+  \dot\mfc_t\right)
= e^{\alpha_t}\left(-\tfrac12\sigma^2_t\mfc^2_t + \zeta_t\right)
=-\tfrac12 e^{-\alpha_t}\sigma^2_t\, (e^{\alpha_t}\mfc_t)^2 + e^{\alpha_t}\,\zeta_t,
\]
where
\[
\zeta_t:=\frac{1}{2\sigma^2_t}\sum_{k\in\mcK}\pi_k\,\big(\beta_t^{(k)}\big)^2.
\]
Let $\tilde{\mfc_t}:=e^{\alpha_t}\,\mfc_t$, then we have
\[
\tfrac{\d}{\dt}(\tilde\mfc_t)
=-\tfrac12 e^{-\alpha_t}\sigma^2_t\, \tilde\mfc_t^2 + e^{\alpha_t}\,\zeta_t
\]
Defining, $\tilde\sigma_t:=\tfrac12 e^{-\alpha_t}\sigma^2_t$ and $\tilde\zeta_t:=e^{\alpha_t}\,\zeta_t$, we may re-write the above as
\[
\tfrac{\d}{\dt}(\tilde\mfc_t) = -\tilde\sigma_t\, \tilde\mfc_t^2 + \tilde\zeta_t.
\]
Note, the coefficient of $\tilde\mfc_t^2$ on the rhs is negative, consequently,
\begin{align}
    \tfrac{\d}{\dt}(\tilde\mfc_t) \le \tilde\zeta_t
    \qquad \Rightarrow \qquad 
    \tilde\mfc_T-\tilde\mfc_t \le \int_t^T \tilde\zeta_s\,\ds
    \qquad \Rightarrow \qquad
    \tilde\mfc_t \ge -\int_t^T \tilde\zeta_s\,\ds
    \label{eqn:mfc-lower-bound}
\end{align}
This provides a lower bound on $\tilde{\mfc}_t$ and hence a lower bound of $\mfc_t$.
Next, we obtain an upper bound. By \eqref{eqn:c-ode}, we deduce that $\left.\left(\frac{\d}{\dt}\tilde\mfc_t\right)\right|_{t=T} = \zeta_T>0$, hence there exists $\ep>0$, s.t. $\forall t\in[T-\ep,T)$, $\tilde\mfc_t< 0$.

Next, take $s=\sup\{u\in[0,T): \tilde\mfc_u=0\}$. Then, by \eqref{eqn:c-ode}, we have $\left.\left(\frac{\d}{\dt}\tilde\mfc_t\right)\right|_{t=s} = \zeta_s>0$, and hence there exists $\ep_s>0$, s.t. $\forall u\in[s-\ep_s,s)$, $\tilde\mfc_u< 0$. Repeating this argument for all times $(s_i)_{i\in\mathbb{N}}$ s.t. $\tilde\mfc_{s_i}$ is zero, we conclude that $\tilde\mfc_t\le0$ for all $t\in[0,T]$. Combining with the inequality \eqref{eqn:mfc-lower-bound}, we have
\[
 -\int_t^T \tilde\zeta_s\,\ds\le\tilde\mfc_t\le0,
\]
and therefore we arrive at the bound
\begin{equation}
 -e^{-\alpha_t}\int_t^T \tilde\zeta_s\,\ds\le\mfc_t\le0.
\end{equation}

As $\mfc_t$ remains finite on the interval $t\in[0,T]$,  we may solve for $\mfb_t$ by using an integrating factor as follows:
\begin{align}
    \frac{d}{dt}\left(e^{-\int_t^T (\sigma^2_u\,\mfc_u-\ob_u)\,\du}\,\mfb_t\right)
    &= 
    \left((\sigma^2_t\,\mfc_t-\ob_t)\,\,\mfb_t
    + \dot\mfb_t\right) e^{-\int_t^T (\sigma^2_u\mfc_u-\ob_u)\,\du}
    \\
    &= -\left(\oa_t\,\mfc_t  
    +\frac{1}{\sigma^2_t}
    \sum_{k\in\mcK}\pi_k \,\alpha_t^{(k)}\,\beta_t^{(k)}\right)e^{-\int_t^T (\sigma^2_u\mfc_u-\ob_u)\,\du}\,.
\end{align}
Integrating from $t$ to $T$, and as $\mfb_T=0$, we obtain
\begin{equation}
   -e^{-\int_t^T (\sigma^2_u\,\mfc_u-\ob_u)\,\du}\,\mfb_t  
   = 
   -\int_t^T \left(\oa_s\,\mfc_s  
    +\frac{1}{\sigma^2_s}
    \sum_{k\in\mcK}\pi_k \,\alpha_s^{(k)}\,\beta_s^{(k)}\right)e^{-\int_s^T (\sigma^2_u\mfc_u-\ob_u)\,\du}\,
    \ds
\end{equation}
which implies
\begin{equation}
    \mfb_t  
   = 
   \int_t^T \left(\oa_s\,\mfc_s  
    +\frac{1}{\sigma^2_s}
    \sum_{k\in\mcK}\pi_k \,\alpha_s^{(k)}\,\beta_s^{(k)}\right)e^{\int_t^s (\sigma^2_u\mfc_u-\ob_u)\,\du}\,
    \ds.
\end{equation}
Then for $\mfa_t$, integrating from $t$ to $T$, and as $\mfa_T=0$, we obtain
\begin{equation}
    \mfa_t = \int_t^T
    \left(
     \oa_s\,\mfb_s + \tfrac12\sigma^2_s(\mfc_s+\mfb_s^2) - \frac{1}{2\sigma^2_s}\sum_{k\in\mcK}\pi_k\,\big(\alpha^{(k)}_s\big)^2
     \right)\,\ds\,.
\end{equation}

Due to the boundedness on the coefficients $(a^{(k)},b^{(k)})_{k\in\mcK}$ and the established bound on $\mfc$, we see that $\mfa$ and $\mfb$ are also bounded on the interval $[0,T]$.

Therefore, $L_0(t,x)=-(\mfa_t+\mfb_t\,x+\frac12\mfc_t\,x^2)$ and from \eqref{eqn:theta-0} we obtain the stated result.
\end{proof}

The next proposition establishes that this class of OU expert models satisfy the requirements of \Cref{thm:barycentre}.
\begin{proposition}
    For the expert models specified above, we have that $L_0\in\C^{1,2}([0,T)\times\R^d;\R)\cap
\C^0([0,T]\times\R^d;\R)$ and has at most quadratic growth, and
\[
\E^{\Q[\bmu]}\left[e^{\frac12\int_0^T|\blambda_{0,u}|^2\du}\right]<+\infty\,, 
\quad \E^{\Q[\theta_0]}\left[\sup_{t\in[0,T]}|X_t|^2\right]<+\infty\,,
\]
where $\blambda_{0,t}:=\sigma^{-1}_t\big(\bmu_t-\theta_{0,t}\big)$
and \begin{equation}
    \theta_{0,t} := \bmu_t-\Sigma_t\nabla_x L_0(t,X_t)\,.
\end{equation}
\end{proposition}
\begin{proof}
    First, form \Cref{prop:OU}, we see that $L_0$ satisfies the properties in the proposition statement.
    Next, there exists $C$, s.t.,
    \begin{equation}
        \E^{\Q[\theta_0]}\left[\sup_{t\in[0,T]}|X_t|^2\right] 
        \le C\,\E^{\Q[\theta_0]}\left[\int_0^T d[X,X]_t\right] =
        C\,\E^{\Q[\theta_0]}\left[\int_0^T \sigma_t^2\,\dt\right]< \infty\,.
    \end{equation}
    Next, by \Cref{prop:OU}, $\blambda_{0,t} = \sigma_t \nabla_x L_0(t,X_t) = -\sigma_t(\mfb_t+\mfc_t\,X_t)$, therefore,
    \begin{align}
        \E^{\Q[\bmu]}\left[e^{\frac12\int_0^T|\blambda_{0,u}|^2\du}\right]
        &=
        \E^{\Q[\bmu]}\left[e^{\frac12\int_0^T\sigma_u^2(\mfb_u+\mfc_u\,X_u)^2\du}\right]\,.
    \end{align}
    Define $h(t,x):=\E^{\Q[\bmu]}_{t,x}\left[e^{\frac12\int_t^T\sigma_u^2(\mfb_u+\mfc_u\,X_u)^2\du}\right]$, then $h$ satisfies the PDE
    \begin{equation}
        \left\{
        \begin{array}{rr}
             \partial_t h + \bmu\,\partial_x h + \tfrac12 \sigma^2\,\partial_{xx}h + \frac12 \sigma^2(\mfb + \mfc x)^2&=0\,,  \\
             h(T,x) &= 1\,. 
        \end{array}
        \right.
    \end{equation}
    As this is an affine PDE, it admits a solution of the form $h(t,x)=e^{\mfl_t+\mff_t\,x+\mfg_t\,x^2}$, where $\mfl$, $\mff$, and $\mfg$ are all deterministic functions of time. As $\bmu$, $\sigma$, $\mfb$, and $\mfc$ are all bounded, using similar arguments as in the proof of \Cref{prop:OU}, we find that $\mfl$, $\mff$, and $\mfg$ are all bounded for all $t\in[0,T]$. Therefore, we conclude $\E^{\Q[\bmu]}\left[e^{\frac12\int_0^T|\blambda_{0,u}|^2\du}\right]=h(T,x) <\infty$.
\end{proof}

\section{The case with beliefs}\label{sec:solution}

In this section, we extend the barycentre discussed above to the case when the agent has additional beliefs that they wish to incorporate. That is, we solve \eqref{opt:P}.

\subsection{Optimal model}
Beliefs may be incompatible with the experts' models. To avoid this scenario, we make the following assumption, which guarantees that the constraints imposed by the agent are feasible.
\begin{assumption}[Feasibility of constraints]\label{asm:eta-star}
 A solution to the following set of equations exists
\begin{equation}
\label{eq:eta-solution}
    \nabla_\eta \log \left(\E^{\Q^\bary}\left[e^{-\eta_0\int_0^T g_t\,\dt - \eta_1\,f_T}\right]\right) = 0\,.
\end{equation}
\end{assumption}

\begin{lemma}
    If the random variable $Y:=(\int_0^T g_t\,\dt,f_T)^\intercal$ is $\Q^\bary$--non-degenerate:
    \[
    \forall c\in\R^2,\qquad 
    \Q^\bary\left(\{\omega:Y(\omega)=c\}\right)<1,
    \]
    i.e., $Y$ is $\Q^\bary$--a.s. not constant, then if there exists a solution to \eqref{eq:eta-solution}, it is unique and we denote this unique solution by $\eta^*:=(\eta^*_0, \eta^*_1)$.
\end{lemma}
\begin{proof}
    Let $K(\eta):=\E^{\Q^\bary}[e^{-\eta\cdot Y}]$. We first show that $K(\eta)$ is strictly convex. To this end, note that
    \begin{align}
        \nabla_\eta K(\eta) = -\frac{\E^{\Q^\bary}[Y\, e^{-\eta\cdot Y}]}{\E^{\Q^\bary}[e^{-\eta\cdot Y}]}
         = - \E^{{\Q^\bary_\eta}}\left[Y\right]
    \end{align}
    and
    \begin{align}
        \nabla^2_{\eta\eta} K(\eta) &= \frac{\E^{\Q^\bary}[Y\,Y^\intercal\, e^{-\eta\cdot Y}]}{\E^{\Q^\bary}[e^{-\eta\cdot Y}]}
        - \frac{\E^{\Q^\bary}[Y\, e^{-\eta\cdot Y}]\E^{\Q^\bary}[Y^\intercal\, e^{-\eta\cdot Y}]}{\left(\E^{\Q^\bary}[e^{-\eta\cdot Y}]\right)^2}
        \nonumber
        \\
        &= \E^{{\Q^\bary_\eta}}[Y\,Y^\intercal] -  \E^{{\Q^\bary_\eta}}[Y] \,\E^{{\Q^\bary_\eta}}[Y^\intercal]
        \\
        &=\Cov^{{\Q^\bary_\eta}}[Y]
        \label{eqn:hess-K}
    \end{align}    
    where the measure ${\Q^\bary_\eta}$ is defined by the Radon-Nikodym derivative
    \[
    \frac{\dQ_\eta}{\dQ^\bary} = \frac{e^{-\eta\cdot Y}}{\E^{\Q^\bary}[e^{-\eta\cdot Y}]}\,,
    \]
    and $\Cov^{{\Q^\bary_\eta}}[Y]$ denotes the ${\Q^\bary_\eta}$-covariance matrix of $Y$.
    As $Y$ is ${\Q^\bary_\eta}$--non-degenerate and as ${\Q^\bary_\eta}$ and $\Q^\bary$ are equivalent, it is also $\Q^\bary_\eta$--non-degenerate. Hence, $\Cov^{{\Q^\bary_\eta}}[Y]$ and therefore $\nabla_{\eta\eta}^2K(\eta)$, is strictly positive definite. Therefore $K(\eta)$ is strictly convex. Consequently, if a minimiser of $K$ exists, it is unique, and solves the first order condition $\nabla_\eta K(\eta)=0$. Thus, there can be at most one solution to \eqref{eq:eta-solution}.
\end{proof}

The theorem below provides our next key result the solution to optimisation problem \eqref{opt:P}, i.e., the drift under the optimal measure, given in \eqref{eqn:theta-eta}, and the associated value function, given in \eqref{eqn:L-eta}.
\begin{theorem}[Optimal Drift and Value Function with Beliefs]
\label{thm:value-beliefs}
Let Assumptions \ref{asm:strong=sol-SDE} and \ref{asm:compatibility}  be satisfied. 
Take $\eta\in\R^2$ such that $\E^{\Q^\bary}[\exp\{-\eta_0\int_0^T g_t\,\dt - \eta_1\,f_T\}]<+\infty$. 

Define the function $L_\eta:[0,T]\times\R^d\to\R$, s.t.
\begin{equation}
\label{eqn:L-eta}
    L_\eta(t,x) := -\log \E^{\Q[\bmu]}_{t,x}\left[ 
    \; e^{- \int_t^T (\varsigma(u,X_u)\, + \eta_0\, g_u)\, \du - \eta_1\,f_T } \; 
    \right],
    \qquad \forall (t,x) \in [0,T]\times\R^d\,.
\end{equation}
Suppose that  $L_\eta\in\C^{1,2}([0,T)\times\R^d;\R)\cap
\C^0([0,T]\times\R^d;\R)$  and has at most quadratic growth, i.e. there exists $C\in\R_+$ s.t. $|L_\eta(t,x)|\le C(1+|x|^2)$ for all $(t,x)\in[0,T]\times\R^d$. Next, define the process $\theta_\eta:=(\theta_{\eta,t})_{\tT}$ 
\begin{equation}
\label{eqn:theta-eta}
    \theta_{\eta,t} := \bmu_t-\Sigma_t\nabla_{x}L_\eta(t,X_t)\,,
\end{equation}
and suppose that
\begin{align*}
&\E^{\Q[\bmu]}\left[e^{\frac12\int_0^T|\blambda_{\eta,u}|^2\du}\right]<+\infty\,, 
\quad \E^{\Q[\theta_\eta]}\left[\sup_{t\in[0,T]}|X_t|^2\right]<+\infty\,,
\\[0.5em]
&\E^{\Q[\theta_\eta]}\left[|f_T|\right]<+\infty\,,
\quad \text{and} \quad
\E^{\Q[\theta_\eta]}\left[\int_0^T|g_u|\du\right]<+\infty\,,
\end{align*}
where $\blambda_{\eta,t}:=\sigma^{-1}_t\big(\bmu_t-\theta_{\eta,t}\big)$.
Then $\theta_\eta$ is admissible, $L_\eta$ is the value function in \eqref{eq:optimal-performance}, and the probability measure that attains the minimum is $\Q[\theta_\eta]$.
\end{theorem}
\begin{proof}
The proof follows along the lines of \Cref{thm:barycentre} and is omitted for brevity.
\end{proof}

Similarly to the pure barycentre case, we have the following representation for the RN derivative of the optimal measure.
\begin{proposition}[Measure change representation]
\label{prop:belif-RN}
Under the assumptions of \Cref{thm:value-beliefs}, and for $\eta \in \R^2$, the RN derivative $\frac{\d \Q[\theta_\eta]}{\dQ[\bmu]} $ has representation
\begin{equation}
\label{eqn:belief-RN}
    \frac{\d \Q[\theta_\eta]}{\dQ[\bmu]} 
    = 
    \frac{e^{- \int_0^T (\varsigma(t,X_t)+\eta_0\,g_t\,\dt - \eta_1\, f_T}}{\E^{\Q[\bmu]}\left[
    e^{- \int_0^T (\varsigma(t,X_t)+\eta_0\,g_t\,\dt - \eta_1\, f_T}\right]}\;.
\end{equation}
\end{proposition}
\begin{proof}
    The proof follows along the same lines as \Cref{prop:bary}, and is omitted for brevity.
\end{proof}

Thus far, we have the value function, drift, and optimal measure identified for problem \eqref{eq:optimal-performance} for an arbitrary $\eta$. To obtain the solution to orignal problem \eqref{opt:P}, the optimal Lagrange multipliers $\eta^\star$ must be chosen to bind the constraints, that is to solve
\begin{equation*}
    \E^{\Q[\theta_{\eta^\star}]}\Big[\int_0^T g_u\, \du\Big]=0 
    \qquad \text{and} \qquad
    \E^{\Q[\theta_{\eta^\star}]}\big[f_T\big]=0\,\,.
\end{equation*}

The following result shows that the probability measure that solves \eqref{opt:P} exists and is unique. Moreover, it has a RN derivative, relative to the average drift measure, of the exponential type, similar to the barycentre measure.
\begin{proposition}[Optimal change of measure]\label{prop:RN-representation}
Let the assumptions in \Cref{thm:value-beliefs} and  \Cref{asm:eta-star} hold. Then, there exists a unique solution to optimisation problem \eqref{opt:P}. 
Moreover, the measure that attains the infimum has RN derivative given by
\begin{equation}
\label{eq:dQeta-dQbmu}
    \frac{\dQ[\theta_{\eta^*}]}{\dQ[\bmu]} =
    \frac{e^{- \int_0^T (\varsigma(t,X_t) + \eta^*_0\, g_t)\,\dt - \eta_1^*\,f_T)} }{\E^{^{\Q[\bmu]}}\left[
    e^{- \int_0^T (\varsigma(t,X_t)+ \eta_0^*\, g_t)\,\dt - \eta_1^*\,f_T}\right]}\;,
\end{equation}
where $\eta^*$ solves \eqref{eq:eta-solution} in  \Cref{asm:eta-star}.
\end{proposition}
\begin{proof}
For any $\eta$ the solution to \eqref{eq:optimal-performance} is $\Q[\theta_\eta]$ and given by \eqref{eqn:belief-RN}. Thus, we only need to find the Lagrange multipliers $\eta$ that bind the constraints, which then gives the solution to optimisation problem \eqref{opt:P}. Uniqueness follows by strict convexity of the KL divergence and as the constraints are linear functionals of the probability measure.

For $\eta$ fixed, \Cref{prop:belif-RN} states that
\begin{equation}\label{eq:dQ_eta-dQ-bmu}
    \frac{\dQ[\theta_{\eta}]}{\dQ[\bmu]}
    =
    \frac{e^{- \int_0^T (\varsigma(t,X_t) + \eta_0\, g_t)\,\dt - \eta_1\,f_T} }{\E^{^{\Q[\bmu]}}\left[\, 
    e^{- \int_0^T( \varsigma(t,X_t)+ \eta_0\, g_t)\,\dt - \eta_1\,f_T}\right]}\,.
\end{equation}
Next, using \Cref{prop:bary}, we rewrite this RN derivative as follows
\begin{subequations}\label{eqs:RN derivative-eq}
    \begin{align}
    \frac{\dQ[\theta_{\eta}]}{\dQ[\bmu]}
    &=
    \frac{e^{- \int_0^T (\varsigma(t,X_t) + \eta_0\, g_t)\,\dt - \eta_1\,f_T} }{\E^{^{\Q^\bary}}\left[\, \frac{\dQ[\bmu]}{\dQ^\bary}\, 
    e^{- \int_0^T (\varsigma(t,X_t)+ \eta_0\, g_t)\,\dt - \eta_1\,f_T}\right]}
    \\
    &= 
    \frac{e^{- \int_0^T (\varsigma(t,X_t) + \eta_0\, g_t))\,\dt - \eta_1\,f_T} }{\E^{\Q[\bmu]}\left[e^{-\int_0^T \varsigma(t,X_t)\,\dt }\right]\, \E^{^{\Q^\bary}}\left[\,  
    e^{-  \eta_0\int_0^T g_t\,\dt - \eta_1\,f_T}\right]}
    \\
    &=
    \frac{\d \Q^\bary}{\dQ[\bmu]} \, 
    \frac{e^{- \eta_0\int_0^T  g_t\,\dt - \eta_1\,f_T} }{\E^{^{\Q^\bary}}\left[\,  
    e^{- \eta_0\int_0^T  g_t\,\dt - \eta_1\,f_T}\right]}
    \,. 
\end{align}
\end{subequations}
Continuing, using the above representation of the RN derivative, we rewrite the running constraint 
\begin{align}
\notag
    \E^{\Q[\theta_\eta]}\left[\int_0^T g_u\, \du\right] 
    &=
    \E^{\Q^\bary}\left[ \frac{\dQ[\theta_\eta]}{\dQ[\bmu]}\frac{\dQ[\bmu]}{\dQ^\bary}\int_0^T g_u\, \du\right]
    \\
    &=
    \E^{\Q^\bary}\left[
    \frac{e^{-\eta_0 \int_0^T  g_t\,\dt - \eta_1\,f_T} }{\E^{^{\Q^\bary}}\left[
    e^{- \eta_0\int_0^T  g_t\,\dt - \eta_1\,f_T}\right]}
    \int_0^T g_u\,\du\, \right]
    \notag
    \\
    &=- \frac{\partial}{\partial a}\log\left(\E^{\Q^\bary}\left[
    e^{- a\int_0^T g_t\,\dt - \eta_1\,f_T} \, \right]\right)\Big|_{a = \eta_0}
    \label{eq:const-eta_0}
    \,.
    \end{align}
Similarly, we can rewrite the second constraint as
    \begin{align}
    \label{eq:const-eta_1}
    \E^{\Q[\theta_\eta]}\left[f_T\, \right] 
    &=
    - \frac{\partial}{\partial a}\log\left(\E^{\Q^\bary}\left[
    e^{- \eta_0\int_0^T g_t\,\dt - a\,f_T} \, \right]\right)\Big|_{a = \eta_1}
    \,.
\end{align}
By \Cref{asm:eta-star}, when $\eta = \eta^*$ the Equations \eqref{eq:const-eta_0} and \eqref{eq:const-eta_1} vanish and hence, under the measure $\Q[\theta_{\eta^*}]$ the constraints are binding. Replacing $\eta$ with $\eta^*$ in \eqref{eq:dQ_eta-dQ-bmu} yields the representation of the RN derivative $\frac{\dQ[\theta_{\eta^*}]}{\dQ[\bmu]}$ in the statement of the proposition.
\end{proof}

\subsection{Distorting the barycentre model}
We can view the optimal measure as a distortion of the barycentre measure by incorporating the constraints. Specifically, alternatively to optimisation problem \eqref{opt:P}, we can in a first step find the barycentre and then in a second step distort the barycentre model to incorporate the constraints. 
\Cref{fig:diagram-commute} provides a visualisation of this idea. The left panel shows that we may obtain the barycentre model and then impose the constraints from the single barycentre model or  directly find the optimal model by searching within the set of constraints and minimizing the weighted KL divergence.
\begin{figure}[H]

\centering
\begin{minipage}{0.35\textwidth}
\centering
\begin{tikzpicture}[    mid arrow/.style={
        postaction={decorate,decoration={
            markings,
            mark=at position .8 with {\arrow{Latex[length=6.4pt, sep=-3.2pt]}}
    }}
  },]

\node[measure] (P1) at (0.5,-1.5) {};
\node[] () at (0.25,-1.5) {${\mathcolor{blue}{\P_1}}$};
\node[measure] (P2) at (-2,0) {};
\node[] () at (-2.35,0) {${\mathcolor{blue}{\P_2}}$};
\node[measure] (P3) at (-1.5,1.25) {};
\node[] () at (-1.85,1.25) {${\mathcolor{blue}{\P_3}}$};
\node[measure] (P4) at (1,0.75) {};
\node[] () at (0.65,1) {${\mathcolor{blue}{\P_4}}$};

\node[measure,fill=green!10] (Q) at (-0.4,-0.4) {};
\node[] () at (-.9,-0.25) {$\mathcolor{red}{\Q^\bary}$};
\draw[mid arrow,dashed,color=blue] (P1) to [out=90,in=-25] (Q);
\draw[mid arrow,dashed,color=blue] (P2) to [out=-25,in=245] (Q);
\draw[mid arrow,dashed,color=blue] (P3) to [out=-25,in=90] (Q);
\draw[mid arrow,dashed,color=blue] (P4) to [out=225,in=45] (Q);

\fill[green!10] (1,-2) -- (0.75, -1) -- plot[domain=0.5:2.5] (\x,{0.25*sin(0.5*\x*pi r)}) -- (2.5,-2) -- cycle;
\node[] () at (2,-1.5) {$\mathcal{C}$};

\node[measure,fill=green!10] (QC) at (0.9,-0.25) {};
\node[] () at (1.3,-0.45) {$\mathcolor{red}{\Q^*}$};

\draw[mid arrow,dashed,color=red] (Q) to [out=0,in=180] (QC);

\end{tikzpicture}

\end{minipage}
\begin{minipage}{0.35\textwidth}
\centering
\begin{tikzpicture}[ mid arrow/.style={
        postaction={decorate,decoration={
            markings,
            mark=at position .8 with {\arrow{Latex[length=6.4pt, sep=-3.2pt]}}
    }}
  },]
\node[measure] (P1) at (0.5,-1.5) {};
\node[] () at (0.25,-1.5) {${\mathcolor{blue}{\P_1}}$};
\node[measure] (P2) at (-2,0) {};
\node[] () at (-2.35,0) {${\mathcolor{blue}{\P_2}}$};
\node[measure] (P3) at (-1.5,1.25) {};
\node[] () at (-1.85,1.25) {${\mathcolor{blue}{\P_3}}$};
\node[measure] (P4) at (1,0.75) {};
\node[] () at (0.65,1) {${\mathcolor{blue}{\P_4}}$};

\fill[green!10] (1,-2) -- (0.75, -1) -- plot[domain=0.5:2.5] (\x,{0.25*sin(0.5*\x*pi r)}) -- (2.5,-2) -- cycle;
\node[] () at (2,-1.5) {$\mathcal{C}$};

\node[measure,fill=green!10] (QC) at (0.9,-0.25) {};
\node[] () at (1.3,-0.45) {$\mathcolor{red}{\Q^*}$};
\draw[mid arrow ,dashed,color=blue] (P1) to [out=0,in=-90] (QC);
\draw[mid arrow ,dashed,color=blue] (P2) to [out=-25,in=135] (QC);
\draw[mid arrow ,dashed,color=blue] (P3) to [out=-25,in=90] (QC);
\draw[mid arrow ,dashed,color=blue] (P4) to [out=0,in=45] (QC);

\end{tikzpicture}
\end{minipage}
\caption{Illustration that first finding the barycenter model and imposing constraints is equivalent to directly imposing the constraints and minimizing the weighted KL divergence.}
\label{fig:diagram-commute}
\end{figure}

This idea is formalised in the next statement. Modifying stochastic processes to include expectation and running cost constraints has been studied in \cite{Jaimungal2024SICON} for L\'evy--It\^o processes, and in \cite{Kroell2024IME} for pure jump processes in a financial risk management setting.
\begin{proposition}[Alternative optimisation problem]\label{prop:equivalent-opt}
Let the assumptions in \Cref{thm:value-beliefs} and \Cref{asm:eta-star} hold. Then, optimisation problem \eqref{opt:P} is equivalent to 
    \begin{align}\label{opt:prime}
    \tag{$P^\prime$}
    \inf_{\Q\in\mcQ^\bary}   \,\mcH\left[\Q\,\big\| \,\Q^\bary\right], \qquad s.t.\quad &\E^\Q\left[\int_0^T g(u, X_u)\, \du\,\right] = 0\,, \quad \text{and} \quad 
    \\[1em]
    & \E^\Q[f(X_T)]=0\,,
    \notag
\end{align}
where
\begin{align*}
    \mcQ^\bary:= \Bigg\{\; \Q[\theta]~\Big|~
     & 
     \frac{\dQ[\theta]}{\dQ^{\bary}} = \exp\left\{-\tfrac{1}{2}\int_0^T |\lambda_t^\bary|^2\dt-\int_0^T \lambda_t^{\bary\intercal}\,\dW_t^{\bary}\right\}\,,
     \\[0.5em]
     &
      \text{where} \quad \lambda_t^\bary:=\sigma^{-1}_t\big(\mu^{\bary}_t-\theta_t\big)\,,
    \\[0.5em]
    & \quad \theta:=(\theta_t)_{\tT} \text{ is an $\F$-adapted process,}
    \\[0.5em]
    & 
    \qquad      
    \text{and} \quad  \E^{\Q^{\bary}}\left[\frac{\dQ[\theta]}{\dQ^{^\bary}}\right]=1\
     \Bigg\}\,,
\end{align*}
 and $\mu^\bary$ is the drift of the barycentre model and  $W^\bary$ is a $\Q^\bary$-Brownian motion.
\end{proposition}

\begin{proof}
   From Corollary 2.9 in \cite{Jaimungal2024SICON}, the solution to \eqref{opt:prime} has RN derivative given by
    \begin{equation*}
        \frac{\dQ^\dagger}{\dQ^\bary} 
        :=
        \frac{e^{- \eta_0^*\int_0^T   g_u\,\du - \eta_1^*\,f_T} }{\E^{^{\Q^\bary}}\left[
    e^{- \eta_0^*\int_0^T  g_u\,\du - \eta_1^*\,f_T}\right]}\,,
    \end{equation*}
where $\eta^*$ is the solution to Equation \eqref{eq:eta-solution}. 
Recall that the solution to optimisation problem \eqref{opt:P}, $\Q[\theta_{\eta^*}]$, is given by \Cref{prop:RN-representation}. 

Next, using \eqref{eqs:RN derivative-eq} we obtain
\begin{align*}
    \frac{\dQ[\theta_{\eta^*}]}{\dQ^\bary}
    &=
    \frac{\dQ[\theta_{\eta^*}]}{\dQ[\bmu]} 
    \frac{\dQ[\bmu]}{\d \Q^\bary}
    =
    \frac{e^{- \eta_0^*\int_0^T  g_u\,\du - \eta_1^*\,f_T} }{\E^{^{\Q^\bary}}\left[\,  
    e^{- \eta_0^*\int_0^T  g_u\,\du - \eta_1^*\,f_T}\right]}
    \,,
\end{align*}
where the Lagrange multipliers $\eta^*$ are the solution to Equation \eqref{eq:eta-solution}. Therefore, it holds that $ \frac{\dQ[\theta_{\eta^*}]}{\dQ^\bary} = \frac{\dQ^\dagger}{\dQ^\bary} $ almost everywhere, which concludes the proof.
\end{proof}

\begin{remark}[Multiple constraints]
     Optimisation problems \eqref{opt:P} and \eqref{opt:prime} can be generalised to multiple constraints of the form
    \begin{align*}    
     \E^\Q\left[\int_0^T g^i(u, X_u)\, \du\right] &= 0\, ,
    \quad \text{for} \quad i = 1, \ldots, I\,,
    \quad \text{and} \quad
    \\[0.5em]
        \notag
    \E^\Q[f^j(X_T)]&=0\,,
    \quad \text{for} \quad j = 1, \ldots, J\,,
    \end{align*}
where $I,J \in \N$, $g^i\colon \R_+ \times \R^d \to \R$, $i = 1, \ldots, I$ and $f^j \colon \R^d \to \R$, $j = 1, \ldots, J$. 

Moreover, all the results in this section, in particular \Cref{thm:value-beliefs}, \Cref{prop:RN-representation}, and \Cref{prop:equivalent-opt}, hold when replacing $\eta_0 g (\cdot, \cdot)$ with $ \boldsymbol{\eta_0} \cdot \mathfrak{g}(\cdot, \cdot)$, where $\boldsymbol{\eta_0}:= (\eta_0^1, \ldots \eta_0^I)$ and $\mathfrak{g}:= (g^1, \ldots, g^I)$, and similarly when replacing $\eta_1 f (\cdot)$ with $ \boldsymbol{\eta_1} \cdot \mathfrak{f}(\cdot)$, where $\boldsymbol{\eta_1}:= (\eta_1^1, \ldots \eta_1^J)$ and $\mathfrak{f}:= (f^1, \ldots, f^J)$, and where $\cdot$ denotes the dot product.

As an illustration of this extension, \Cref{sec:sim-ex} considers a numerical example with three constraints.
\end{remark}

\section{Deep learning algorithms}\label{sec:algos}
In low dimensions, such as $d=1$ or $2$, we can approximate the solution to optimisation problem \eqref{opt:P} using finite-difference methods by (a) solving for the optimal Lagrange multipliers by solving the appropriate PDE, and (b) once the optimal Lagrange multipliers are obtained, using a second finite-difference scheme to approximate $\omega(t,x)$ which then provides, through \eqref{eqn:theta-eta}, the optimal drift. In higher dimensions, however, finite-difference and other traditional PDE methods run into difficulties due to the curse of dimensionality. Hence, to overcome the curse of dimensionality, we propose two deep learning algorithms to approximate the solution to optimisation problem \eqref{opt:P}. The first deep learning algorithm approximates the drift of a candidate measure and uses the difference in the RN densities as a loss function. The second deep learning algorithm approximates a transformation of the value function, specifically $\omega(t,x)= e^{-L_{\eta^*}(t,x)}$, by leveraging the notion of elicitability --- which allows one to estimate conditional statistics of a distribution function, e.g., a conditional expectation, via a strictly convex optimization problem. 
Often when elicitability is used to estimate conditional statistics, a parametric assumption on the dependence of the conditioning variables are made, e.g.,   the dependence is assumed to be linear, quadratic, or polynomial in the conditioning variables. Rather, when utilizing neural networks, due to their universal approximation ability, parametric assumptions are not necessary.
While neural networks and deep learning approaches have been utilised in a variety in financial modelling contexts, the specific algorithms presented here are tailored to the constrained KL barycentre problem that we pose. In this sense, the algorithms are novel. We make use of the fact that neural networks are universal approximators, but tune the algorithms to providing approximations to our specific problem formulation.

\subsection{Learning the optimal drift}\label{sec:algo-drift}

In this section, we develop a deep learning approach that directly learns the stochastic exponential --- the RN derivative --- that drives the measure change from $\Q[\bmu]$ to the optimal measure $\Q[\theta_{\eta^*}]$. For this, we write the measure change from a candidate measure $\Q[\theta]$ to $\Q[\bmu]$ as 
\begin{equation}
\label{eqn:stoch-exp}
    \frac{\dQ[\theta]}{\dQ[\bmu]} = \exp\left\{-\tfrac{1}{2}\int_0^T |\blambda(t,X_t)|^2\,\dt-\int_0^T \blambda(t,X_t)^\intercal\,\dbW_t\right\}, 
\end{equation}
where $\blambda(t,x):=\sigma^{-1}(t,x)(\bmu(t,x)-\theta(t,x))$ and $(\bW_t)_{\tT}$ are $\Q[\bmu]$-Brownian motions. 
To approximate the optimal measure, we parametrise the drift process (under the candidate measure $\Q[\theta]$)  $(\theta_t)_{\tT}$ by a neural network with parameters $\thetaparam$, and write this as $\theta[\thetaparam](t, X_t)$ --- which takes $t$ and $X_t$ as inputs and outputs the drift at time $t$ and state $X_t$.

The approach we take to learning the drift is as follows. First, introduce a time grid $\{t_0,\dots,t_N\}$, where $t_i:=i\Delta t$, $i = 0, \ldots, N$, and $\Delta t :=\frac{T}{N}$. Second,  simulate sample paths of Brownian motions under the measure $\Q[\bmu]$, denoted  as $\bW_{t_0},\bW_{t_1},\dots,\bW_{t_N}$ and write their increments as $\Delta\bW_{t_i}:=\bW_{t_{i+1}}-\bW_{t_i}$. Third, use an Euler–Maruyama discretisation of $X$, which we denote ${\tilde{X}}$, via
\[
\tilde{X}_{t_i} = \tilde{X}_{t_{i-1}} + \bmu_{t_{i-1}}  \,\Delta t  + \sigma_{t_{i-1}} \,\Delta \bW_{t_i}\,,
\]
where $\bmu_{t_{i-1}}:=\bmu(t_{i-1}, \tilde{X}_{t_{i-1}})$ and $\sigma_{t_{i-1}}:=\sigma(t_{i-1}, \tilde{X}_{t_{i-1}})$ --- with a slight abuse of notation we do not place tildes on $\bmu$ and $\sigma$, despite them being evaluated along the Euler--Maruyama discretisation of $X$. From these sample paths, for the current estimate of the parameters $\thetaparam$, we then obtain sample paths of $\theta[\thetaparam]$ by evaluating the neural network for  $\theta[\thetaparam]$ along the sample paths of $\tilde{X}$ that we already obtained, i.e., by evaluating $\theta_{t_i}^\thetaparam :=\theta[\thetaparam](t_{i},\tilde{X}_{t_i})$. We then evaluate
\begin{align*}
    \blambda_{t_{i-1}} = \sigma_{t_{i-1}} ^{-1}\big(\bmu_{t_{i-1}}-\theta_{t_{i-1}}^\thetaparam\big)\,,
\end{align*} 
for all $i=0,\dots,N$, across all sample paths.

With the sample paths of $X$ under $\Q[\bmu]$, we then obtain the Lagrange multipliers by solving for the root of the equations
\begin{equation}\label{eq:constraints-under-bmu}
    \E^{\Q[\bmu]}\left[\frac{\dQ[\theta_{\eta}]}{\dQ[\bmu]} \,f_T\right] = 0\,
    \quad \text{and}\quad 
    \E^{\Q[\bmu]}\left[\frac{\dQ[\theta_{\eta}]}{\dQ[\bmu]} \,    \int_0^T g_u\,\du\right] = 0\,.
\end{equation}
Note that the above quantities can be approximated from samples under the fixed measure $\Q[\bmu]$. The algorithm for finding $\eta^*$ is presented in \Cref{algo:optima-eta}. Note that while \Cref{asm:eta-star} gives $\eta^*$ as a solution to a set of non-linear equations, the expectations are with respect to the barycentre measure $\Q^\bary$. Thus, to utilise \Cref{asm:eta-star}, we need samples paths under $\Q^\bary$, which require to solve for the barycentre model. The above approach makes use of the readily available average drift model, which is straightforwardly obtained from simulations of the experts' models.

Finally, to learn the optimal drift, we use the loss function
\begin{equation*}
\mfL\big[\theta[\thetaparam]\big]
    := 
    \E^{\Q[\bmu]}\left[
   \left( \log\left(\frac{\dQ \big[\theta[\thetaparam]\big]}{\dQ[\bmu]}\right) 
    -
    \log\left(\frac{\dQ[\theta_{\eta^*}]}{\dQ[\bmu]}\right)\right)^2\right]
 \,,
\end{equation*}
with $\frac{\dQ [\theta[\thetaparam]]}{\dQ[\bmu]}$ given in \eqref{eqn:stoch-exp} and $\frac{\dQ[\theta_{\eta^*}]}{\dQ[\bmu]}$ given in \Cref{prop:RN-representation}.
All terms in the loss function may be approximated by simulating under the single measure $\Q[\bmu]$. We update the parameters $\thetaparam$ using gradient descent, via
$\thetaparam \leftarrow \thetaparam - r_\thetaparam\,\nabla_\thetaparam \mfL\big[\theta[\thetaparam]\big]$, where $r_\thetaparam$ is a learning rate. Further details are given in \Cref{algo:girsanov-II}.

\begin{algorithm}[tbp]
\setcounter{AlgoLine}{0}
\scriptsize
    \SetKwRepeat{Do}{do}{while}
	\caption{Learning Optimal $\eta$}
	\label{algo:optima-eta}
	\SetKwInOut{Input}{Input}
    \Input{$\eta=\epsilon \mathds{1}$ and $tol$;}

    generate batch of paths 
    $( \bW_{t_0}^{\msim},\ldots, \bW_{t_{N}}^{\msim})_{m\in\mfM}$ and $(\tilde{X}_{t_0}^{\msim},\ldots, \tilde{X}_{t_N}^{\msim})_{m\in\mfM}$ from $\Q[\bmu]$;
    
    approx. \eqref{eq:dQ_eta-dQ-bmu} to obtain samples $\left(\left(\frac{\dQ[\theta_{\eta}]}{\dQ[\bmu]}\right)^{\msim}\right)_{m\in\mfM}$;

    approx. the constraint expectations in \eqref{eq:constraints-under-bmu} by
    \[
    \hf[\eta]:=\frac1M\sum_{m\in\mfM} f\big(\tilde{X}_{t_N}^{\msim}\big)\,\left(\frac{\dQ[\theta_{\eta}]}{\dQ[\bmu]}\right)^{\msim}
    \quad \text{and} 
    \quad
    \hg[\eta]:=\frac1M\sum_{\substack{m\in\mfM
    \\ i  = 0, \ldots, N-1}} g\big(t_i, \tilde{X}_{t_i}^{\msim}\big)\,\left(\frac{\dQ[\theta_\eta]}{\dQ[\bmu]}\right)^{\msim}\, \Delta t;
    \]
    
    use root finder to obtain $\eta^*$ s.t. $\Big|\hf[\eta^*]\Big|<tol$ and $\Big|\hg[\eta^*]\Big|<tol$;
\end{algorithm}

\begin{algorithm}[tbp]
\setcounter{AlgoLine}{0}
\scriptsize
    \SetKwRepeat{Do}{do}{while}
	\caption{Learning Optimal Drift}
	\label{algo:girsanov-II}
	\SetKwInOut{Input}{Input}
    \Input{NN parameters $\thetaparam$;}
    \Input{ $r_\thetaparam=0.001$;}

    Use Algorithm \ref{algo:optima-eta} to obtain $\eta^*$;
        
    \Do{not converged}
    {
        generate batch of paths 
        $( \bW_{t_0}^{\msim}\dots \bW_{t_{N}}^{\msim})_{m\in\mfM}$ and $(\tilde{X}_{t_0}^{\msim}\dots \tilde{X}_{t_N}^{\msim})_{m\in\mfM}$ from $\Q[\bmu]$;

        use batch to generate  $(\theta_{t_0}^{\msim}[\thetaparam],\dots,\theta_{t_N}^{\msim}[\thetaparam])_{m\in\mfM} $, and $(\blambda_{t_0}^{\msim}, \dots,\blambda_{t_N}^{\msim})_{m\in\mfM}$;    

        approx. $\int_0^T |\blambda_u|^2\du$ and $\int_0^T \blambda_u^\intercal \dbW$ and use \eqref{eqn:stoch-exp} to generate samples $\Big(\Big(\frac{\dQ[\theta[\thetaparam]]}{\dQ[\bmu]}\big)^{\msim}\Big)_{m\in\mfM}$; 

        approx. \eqref{eq:dQeta-dQbmu} to obtain samples $\Big(\Big(\frac{\dQ[\theta_{\eta^*}]}{\dQ[\bmu]}\Big)^{\msim}\Big)_{m\in\mfM}$;
        
        compute loss
        \[
        \mfL\big[\theta[\thetaparam]\big]:= \frac{1}{M}\sum_{m\in\mfM} \left(\log \left(\frac{\dQ\big[\theta[\thetaparam]\big]}{\dQ[\bmu]}\right)^{\msim}
        - \log\left(\frac{\dQ[\theta_{\eta^*}]}{\dQ[\bmu]}\right)^{\msim}\right)^2;
        \]

        update $\thetaparam \leftarrow \thetaparam - r_\thetaparam\,\nabla_\thetaparam \,\mfL\big[\theta[\thetaparam]\big]$;
    }
    \label{algo:drift}
\end{algorithm}

\subsection{Learning the value function}\label{sec:algo-elicit}

In this approach, we approximate $\omega(t,x)=e^{-L_{\eta^*}(t,x)}$, where $\eta^*$ is obtained via \Cref{algo:optima-eta}, using a neural network approximator.  From \Cref{thm:value-beliefs}, we have that with the optimal Lagrange parameters
\begin{equation*}
    \omega(t,x) = \E^{\Q[\bmu]}_{t,x}\left[ 
    \; e^{- \int_t^T (\varsigma(u,X_u) + \eta_0^*\, g_u)\,\du - \eta_1^*\,f_T } \; 
    \right].
\end{equation*}
We then exploit elicitability of conditional expectations to find the best neural network approximator for $\omega$ denoted $\omega[\omegaparam]$ where $\omegaparam$ are the parameters of the NN. Leveraging elicitability for estimating conditional functionals using deep learning has been used in \cite{Fissler2023EJOR, coache2023conditionally, Jaimungal2023WP} to avoid time consuming nested simulations. We refer the interested reader to \cite{Gneiting2011JASA} for elicitability in a statistical context and Appendix C in \cite{Jaimungal2023WP} for a discussion on conditional elicitability and its application in deep learning. 
For example, estimates of the conditional expectation $\E[Y| \sigma(Z)]$, where $Y$ and $Z$ are $\F$-measurable random variables with $\E[Y]<+\infty$, and $\sigma(Z)$ denotes the sigma algebra generated by $Z$, can be obtained by minimizing the loss function
\[
\mfL := \E\big[(h(Z)-Y)^2\big]
\]
over all functions $h$ s.t., $h(Y)$ is square-integrable. Denoting the optimiser as $h^*$, we would have   $h^*(Z)=\E[Y|\sigma(Z)]$. In our context, conditional elicitability implies that minimising the loss function
\begin{equation}
\mfL\big[\omega[\omegaparam]\big]
    :=
    \argmin_{\omegaparam}\; \E^{\Q[\bmu]}\left[ \int_0^T \left(\omega[\omegaparam](t,X_t) -  e^{- \int_t^T (\varsigma(u,X_u) + \eta_0^*\, g_u)\,\du - \eta_1^*\,f_T }\right)^2\;\dt\right]
    \label{eqn:loss-elicit}
\end{equation}
over neural network parameters  $\mathfrak{o}$ yields a good approximation of $\omega(t,x)$. Indeed by conditional elicitability it holds that
\begin{equation*}
    \omega
    =
    \argmin_{h}\; \E^{\Q[\bmu]}\left[ \int_0^T \left(h(t,X_t) -  e^{- \int_t^T (\varsigma(u,X_u) + \eta_0^*\, g_u)\,\du - \eta_1^*\,f_T }\right)^2\;\dt\right]\,,
\end{equation*}
where the minimum is taken over all square integrable functions $h \colon \R_+ \times \R^d \to \R_+$; for more details see \cite{Gneiting2011JASA} and \cite{Jaimungal2023WP} Appendix C.

With the approximation of $\omega$ at hand, we obtain an estimate of the optimal drift using \eqref{eqn:theta-eta}. Note that $\nabla_x L_\eta (t,X_t)$ can be obtained using backpropagation of $ L_\eta (t,X_t)$ with respect to the input parameters $x$.

In implementation, we use the empirical estimator of the expectation  by generating simulations of $X$ under $\Q[\bmu]$, using the same methodology as in \Cref{sec:algo-drift}, and then approximate the Riemann integrals in \eqref{eqn:loss-elicit} using the same time discretisation as in the simulation of $X$. The detailed steps for estimating $\omega$ using the elicitability methodology are provided in \Cref{algo:elicit}.

\begin{algorithm}[tbp]
\setcounter{AlgoLine}{0}
\scriptsize
    \SetKwRepeat{Do}{do}{while}
	\caption{Learning $\omega$ using Elicitability}
	\label{algo:elicit}
	\SetKwInOut{Input}{Input}
    \Input{NN parameters $\omegaparam$;}
    \Input{ $r_\omegaparam=0.001$;}

    Use Algorithm \ref{algo:optima-eta} to obtain $\eta^*$;
        
    \Do{not converged}
    {
        generate batch of paths 
        $( \bW_{t_0}^{\msim}\dots \bW_{t_{N}}^{\msim})_{m\in\mfM}$ and $(\tilde{X}_{t_0}^{\msim}\dots \tilde{X}_{t_N}^{\msim})_{m\in\mfM}$ from $\Q[\bmu]$;   

        use samples to approximate  
        $(\zeta_{i}^{[m]}:=\big(\int_{t_i}^T \big(\varsigma(u,\tilde{X}_u^{[m]}) + \eta_0^*\,g(u,\tilde{X}_u^{[m]})\big)\,\du\big)_{i\in\{0,\dots,N-1\},m\in\mfM}$; 
        
        compute loss
        \[
        \mfL\big[\omega[\omegaparam]\big]:= \frac{1}{M}\sum_{\substack{m\in\mfM\\ i  = 0, \ldots, N-1}} 
        \left( \omega[\omegaparam](t_i,\tilde{X}_{t_i}^{[m]})
        - e^{-\zeta_i^{[m]}-\eta_1^*\,f(\tilde{X}_{t_N}^{[m]}) }\right)^2 \Delta t;
        \]

        update $\omegaparam \leftarrow \omegaparam - r_\omegaparam\,\nabla_\omegaparam \mfL\big[\omega[\omegaparam]\big]$;
    }
\end{algorithm}
We conduct a comparison case study of the two algorithms in the next section.

\section{Applications}

This section is devoted to illustrations of the proposed methodology of combining experts opinions. \Cref{sec:sim-ex} compares the two deep learning algorithms introduced in \Cref{sec:algos} on simulated examples while \Cref{sec:IV-smile} provides an application to implied volatility (IV) smiles. For the IV smiles application, we combine three different models of IV smiles, estimated from real data, and calculate the minimal weighted KL model that satisfies a constraint on the average skewness of the IV smiles.

\subsection{Simulation examples}\label{sec:sim-ex}

Using the methodology of learning the optimal drift explained in \Cref{sec:algo-drift} and \Cref{algo:girsanov-II}, we ran experiments where we start the training using the same initial neural network, however, the number of steps for the Euler discretisation is increased from 10 to 100 to 1000 steps\footnote{Computation is carried out on an Intel Xeon CPU E5-2630 v4 \@ 2.20GHZ, 10 Core machine equipped with an NVIDIA TITAN RTX GPU (launched in 2018). With a batch--size of 1,024 and 1,000 time steps, our implementation runs 25 iterations per second. Newer machines should find much faster computation times.}.  For simplicity we consider a one-dimensional process with volatility that is constant $\sigma=1$ and where the drifts under the two experts' models are
\begin{align*}
    \mu^{(1)}(t,x) &= 4\,t-0.7\,x\,, \text{ and}
    \\
    \mu^{(2)}(t,x) &= 3\left(t+\sin\left(4\pi\,t+\tfrac1{12}\pi\right)-x\right)\,.
\end{align*}
As both expert models are OU processes with bounded coefficients, the results of  section \Cref{sub-sec:bary-OU} apply.

\begin{figure}
    \centering
    $\vcenter{\hbox{
    \includegraphics[width=0.45\textwidth]{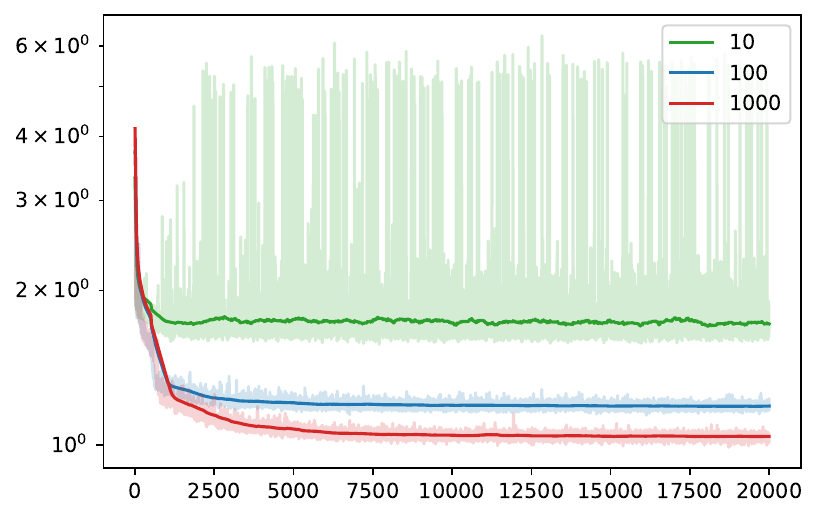}}}$
    $\vcenter{\hbox{
    \includegraphics[width=0.45\textwidth]{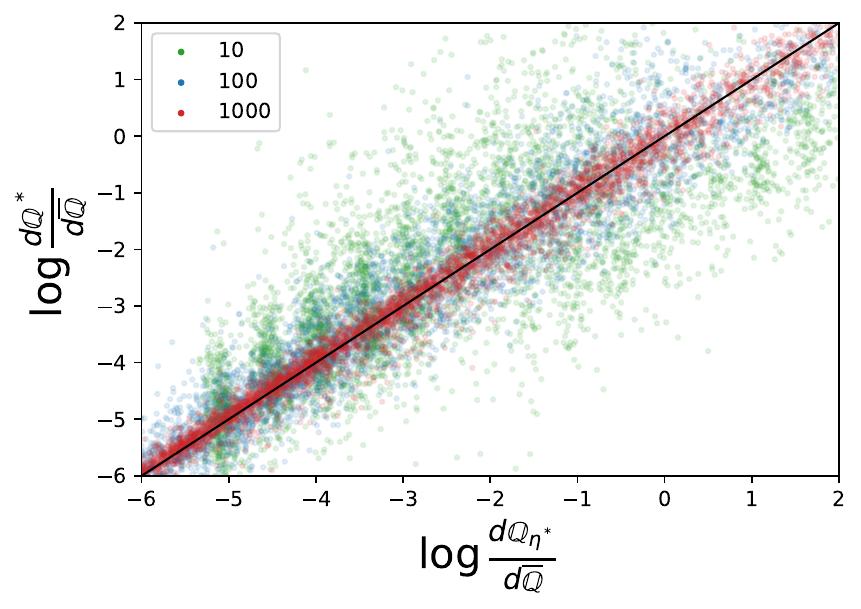}
    }}$
    \\
    \centering
    \includegraphics[width=0.9\textwidth]{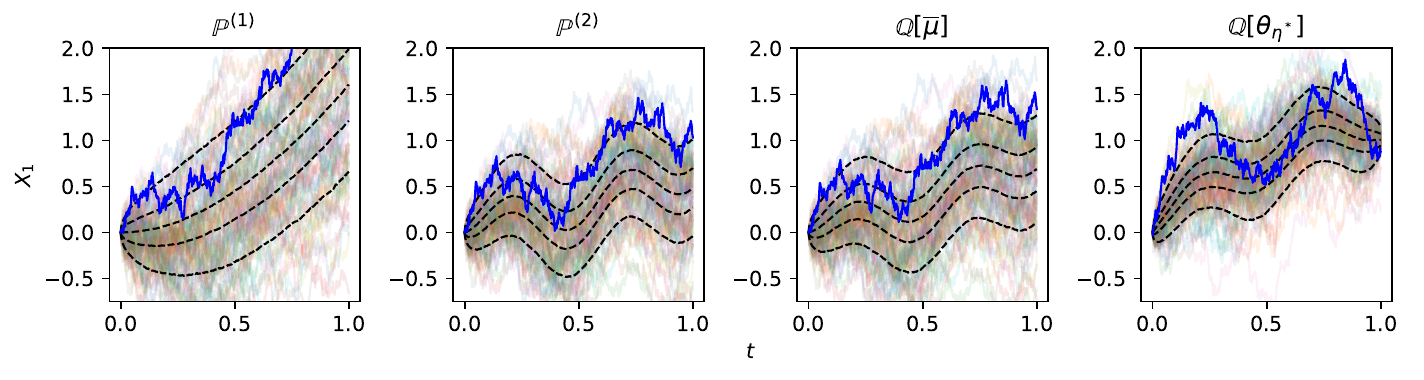}
    \caption{Top: (left) comparison of loss as a function of iteration, and (right) scatter plot of obtain RN derivative versus target, both for various time discretisations. Bottom: evolution under the expert models, average model, and optimal model.}
    \label{fig:loss-scatter-method-2}
\end{figure}

The time horizon is $T = 1$, the experts weights are $\pi_1 =\pi_2 = \frac12$.
Simulated paths under the two expert models are displayed in \Cref{fig:loss-scatter-method-2}. We observe that expert 1's model has an upward drift while expert 2's model is mean-reverting and cyclical. \Cref{fig:loss-scatter-method-2} further displays simulated paths under the average drift model, which combines both the upward drift of expert 1 and the mean-reverting and cyclical pattern of expert 2.

Next, we consider the constraints
\begin{equation*}
    f(x) = \Id_{\{x\in(0.8,1.2)\}}-0.9\,,
    \quad \text{and} \quad
    g(t,x) = \Id_{\{x<t\}}-0.2\,.
\end{equation*}
These choices of constraints mean that the processes at time $T = 1$ should lie with 90\% probability within the interval $(0.8, 1.2)$, that is $\Q(X_1 \in (0.8, 1.2)) = 0.9$, and that the average time the process spends below a barrier is 0.2, specifically $\E^\Q[\int_0^1\Id_{\{X_u<t\}}\du]\allowbreak = 0.2$. Sample paths under the constraint barycentre model are given in the bottom right panel of \Cref{fig:loss-scatter-method-2}. In 
\Cref{tab:constaints} we reports the value of the constraints under the two experts models, the average drift model, and the optimal model $\Q[\theta_{\eta^*}]$, i.e. the neural network approximation of the constrained barycentre. We observe that the mean under the optimal measure for both terminal and running constraints are essentially zero --- in contrast to their expectations under either of the experts' models or the mean drift measure.
\begin{table}[H]
  \centering
  \caption{Constraints under various measures. $\E^\circ$ indicates the expectation under the different probability measures indicated in the columns.}
    \begin{tabular}{lrrrr}
    \toprule
    \toprule
          & \multicolumn{1}{l}{$\P^{(1)}$} & \multicolumn{1}{l}{$\P^{(2)}$} & \multicolumn{1}{l}{$\Q[\bmu]$} & \multicolumn{1}{l}{$\Q[\theta_{\eta^*}]$} \\
          \midrule
    $\E^{\circ}[f(X_1)]$ & $-0.746$ & $-0.717$ & $-0.626$ & $-2.2\times10^{-3}$ \\[0.5em]
    $\E^{\circ}[\int_0^1g(u,X_u)\,\du]$ & $0.285$ & $0.429$ & $0.392$ & 
    $-5.6\times10^{-3}$\\
    \bottomrule\bottomrule
    \end{tabular}%
  \label{tab:constaints}%
\end{table}%
We observe that the value of the constraints of the average drift model is not the average of the value of the constraints of the experts, this is because $\Q[\bmu]$ averages the drifts of both experts' models.

The top two panels of Figure \ref{fig:loss-scatter-method-2} show the corresponding losses during training as well as a scatter plot of the learnt RN derivative $\frac{\d\Q^*}{\d\Q[\bmu]}$, using the representation in \eqref{eqn:stoch-exp},  versus the targeted (true) RN derivative $\frac{\d\Q[\theta_{\eta^*}]}{\d\Q[\bmu]}$ in \eqref{eq:dQeta-dQbmu}
for the three different time discretisations --- using the same underlying Brownian sample paths. If the learnt drift and the simulation are exact, then this scatter plot should be a straight line with slope $1$. When the number of time steps is small, the discrete approximation of the stochastic integrals  cannot capture the optimal measure, generated by $\frac{\d\Q[\theta_{\eta^*}]}{\d\Q[\bmu]}$, path-by-path. This explains why the scatter plot, concentrates more and more along the diagonal (with slope $1$) as the number of time steps increases, and provides strong evidence of the algorithm converges to the correct solution. 
 Further, the bottom panel shows sample paths generated by the same Brownian motions for all four models: $\P^{(1)}$, $\P^{(2)}$, $\Q[\bmu]$, and $\Q[\theta_{\eta^*}]$. We observe that under the optimal measures $\Q[\theta_{\eta^*}]$ at terminal time, the sample paths are concentrated in the interval (0.8, 1.2) and that the sample paths are shifted upwards to enforce both constraints.

Next we examine how \Cref{algo:drift}, the approach where we learn the drift, and \Cref{algo:elicit}, the approach where we learn the value function, compare.
For this, we consider the same underlying models as above, but with the three constraints
\[
f_1(x) = (x-1)\,, \qquad f_2(x)=(x-1)^2-0.05\,, \qquad \text{and}
\qquad g_1(t,x) = \Id_{\left\{x<1-(0.5-t)^2\right\}}-0.8\,.
\]
The first set of constraints functions are constraints on the mean and variance, in particular $\E^\Q[X_1] = 1$ and $\text{var}^\Q(X_1) = 0.05$. The running cost constraint is $\E^\Q[\int_0^1 \Id_{\left\{X_u<1-(0.5-t)^2\right\}}\du] = 0.8$.

\begin{figure}
    \centering
    \includegraphics[width=0.9\textwidth]{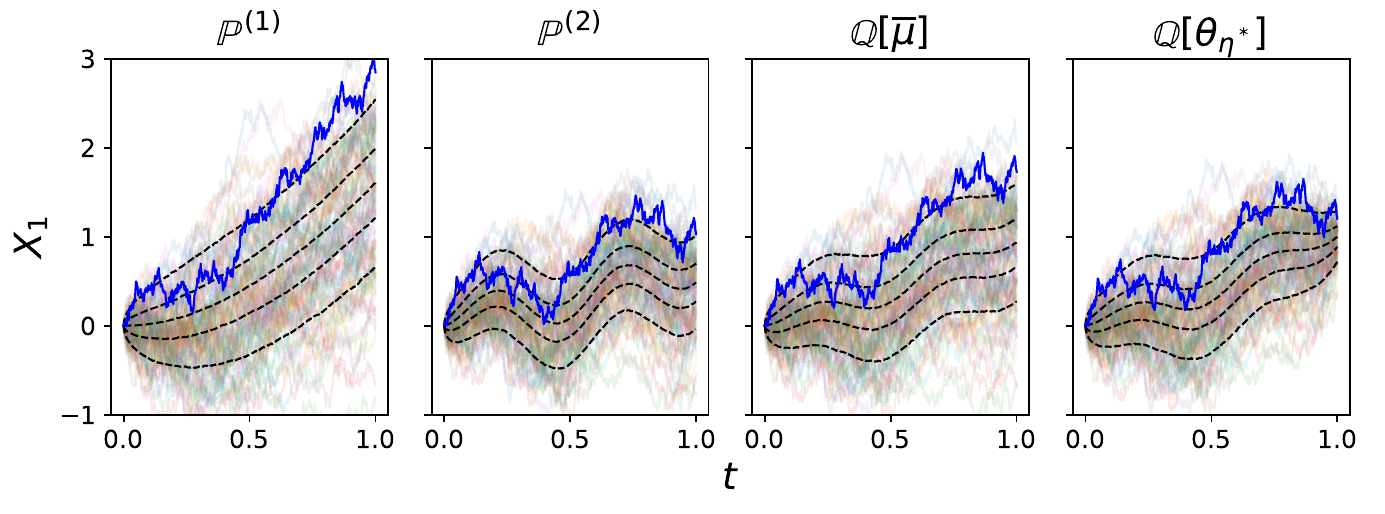}
    \\ 
\includegraphics[width=0.3\textwidth]{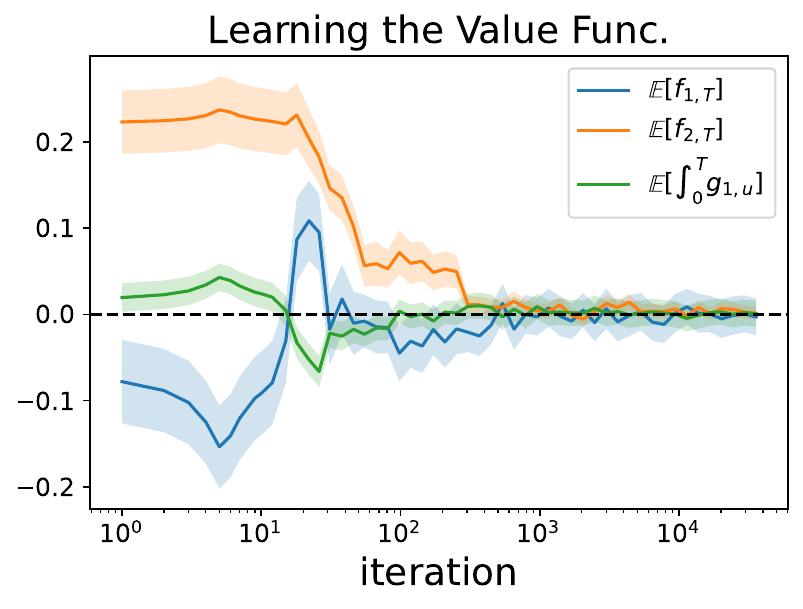}
\includegraphics[width=0.3\textwidth]{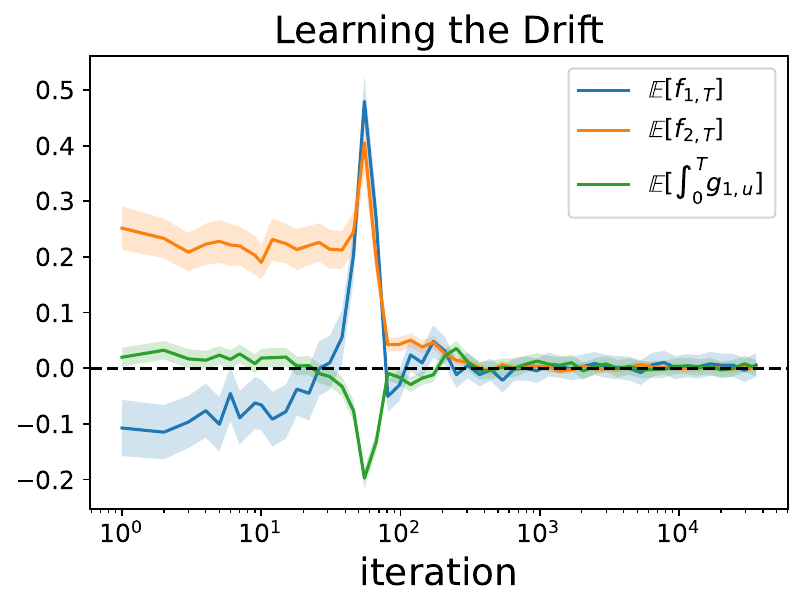}    \includegraphics[width=0.3\linewidth]{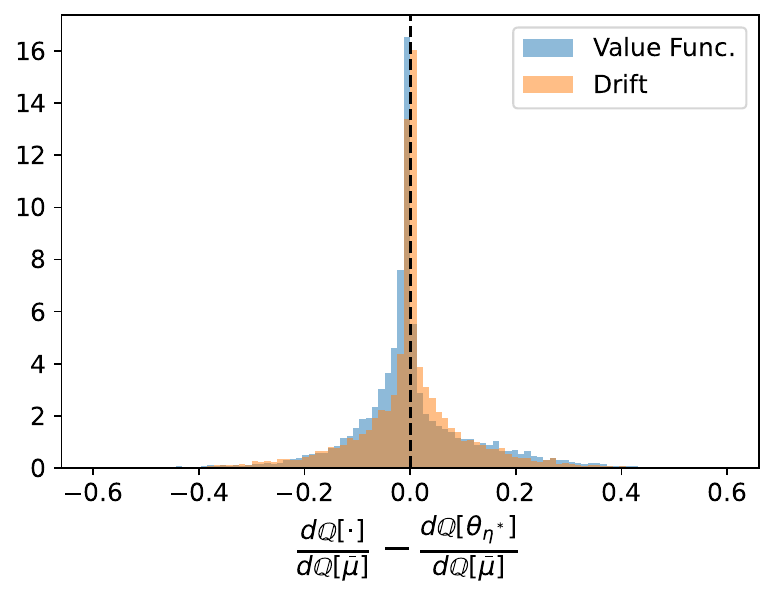}
    \caption{(top) sample paths under the expert models, the average drift model, and the optimal measure. (bottom) The left and middle panels show the evolution of the constraints as the two learning algorithms proceed. The right panel shows the histogram of the difference between the RN derivative of the learnt models minus the target RN derivative $\frac{\dQ[\theta_{\eta^*}]}{\dQ[\bmu]}$.}
    \label{fig:compare-drift-value-func}
\end{figure}

Figure \ref{fig:compare-drift-value-func} shows sample paths of the learnt model in the top panel (both methods produce indistinguishable results) along the two experts models and the average drift model, as well as the evolution of the constraints as training progresses in the bottom panels. The bottom left panel shows the constraints for the ``learning the value function'' algorithm, that is \Cref{algo:elicit} while the bottom middle panel displays the constraints for ``learning the drift'' algorithm, that is \Cref{algo:drift}. After about 1,000 iterations, both approaches lead to constraints that are well within acceptable errors ($\lesssim 1\times 10^{-3}$) --- see the first two panels in the bottom of Figure \ref{fig:compare-drift-value-func}.  The right most figure in the bottom panel shows  the histogram of the difference between the measure change $\frac{\dQ[\theta^*]}{\dQ[\bmu]}$ and $\frac{\dQ[\theta_{\eta^*}]}{\dQ[\bmu]}$ where $\theta^*$ is the drift obtained from either learning the drift (\Cref{algo:drift}) or learning the value function (\Cref{algo:elicit}). They have estimated means of $1.1\times10^{-7}\pm10^{-3}$ and $6.1\times10^{-8}\pm 1.1\times10^{-3}$, respectively. Denoting by $\delta_1$ the difference between the RN derivative obtained by \Cref{algo:drift} and the target RN derivative, i.e. $\delta_1:= 
\frac{\dQ[\theta^*]}{\dQ[\bmu]}-\frac{\dQ[\theta_{\eta^*}]}{\dQ[\bmu]}$, and by $\delta_2$ the difference between the RN derivative obtained by \Cref{algo:elicit} and the target RN derivative, we perform a Welch's T-test for $\delta_1$ and $\delta_2$. The Welch's T-test (which tests for equality of mean in two distributions with differing variances) provides a T-statistic of $-3.26\times10^{-5}$ and corresponding p-value of $0.999974$, indicating we cannot reject the null that they are indistinguishable. Furthermore, the one-sample T-test for each $\delta_1$ and $\delta_2$ individually, with a null hypothesis of a mean of zero, each has T-statistic of $-4.62\times10^{-5}$ and $-2.88\times10^{-6}$, respectively, and corresponding p-values of  $0.99996$ and $0.999998$, respectively. As both of the estimated means are statistically indistinguishable from zero, and one another, this provides confidence that both algorithms converged to the true RN derivative. Finally, we estimate the (constrained) KL divergence \eqref{eqn:KL-in-terms-of-theta}, using Monte Carlo simulations of $10,000$ sample paths, where the optimal $\theta$ stems from  \Cref{algo:drift} or \Cref{algo:elicit}. For \Cref{algo:drift} the estimated (constrained) KL divergence is $1.683\pm0.005$ and for learning the value function, \Cref{algo:elicit}, it is $1.660\pm0.005$. We performed a one-sided Welch's T-Test with null hypothesis that KL divergence from learning the value function has a lower mean than that from learning the drift which resulted in a t-statistic of $-326.61$ and p-value that was (machine) indistinguishable from 0. Hence, we have extremely strong evidence that \Cref{algo:elicit} produces a smaller KL divergence than \Cref{algo:drift}. We anticipate that learning the value function \Cref{algo:elicit} is better than the brute-force approach of learning the drift \Cref{algo:drift}, as the former makes use of the underlying mathematical derivation of the optimal value function, and thus the formula for the optimal drift. 

\subsection{Implied volatility smiles}\label{sec:IV-smile}

\begin{figure}[h]
    \centering
    \includegraphics[width=0.6\textwidth]{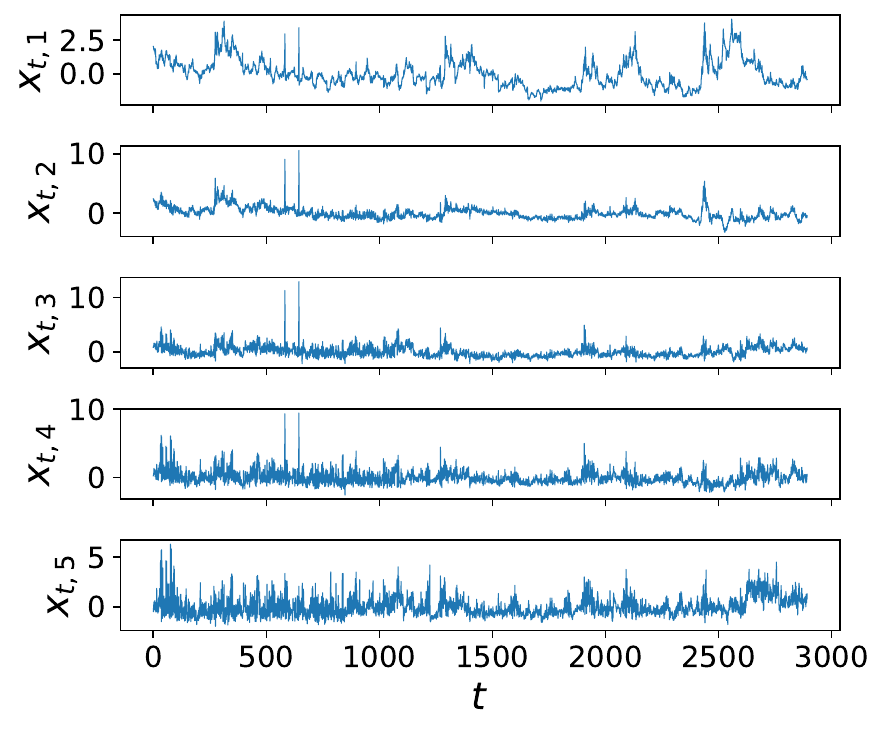}
    \caption{Time series of the (normalised) functional basis coefficients from implied volatility smiles.}
    \label{fig:fda-coeffs}
\end{figure}

In this section we investigate an application to modelling the evolution of implied volatility (IV) smiles --- recall that IV smiles refer to the implied volatility for European options with a fixed time-to-maturity as a function of strike. For this, we estimate three different models
(the expert models) using three distinct time frames and obtain the optimal model with a constraint on the at-the-money skewness. We use daily data for fixed time-to-maturity (TTM) IV smiles from the WRDS database for the assets AMZN at TTM of 60 days over the period July 8, 2010 to December 31, 2021. The data consists of pairs of Delta ($\Delta$) --- rather than strike --- and IV ($\sigma^{IV}$), that is $(\Delta_i,\,\sigma^{IV}_{t,i})_{i=1,\dots,17=:I}$, for days $t = 1, \ldots, T = 2,893$, where $\Delta_i = 0.1,0.15,\dots,0.9$. We approximate the real-world dynamics of the IV smile using the approach in \cite{choudhary2023funvol}, where the authors consider the entire IV surfaces. Here, we restrict to IV smiles for simplicity. In a first stage, for each time $t$, we project  the data onto a functional basis of normalised Legendre polynomials $\{L_j(\Delta)\}_{j=0,\ldots,4=:d-1}$ of up to order $4$ using linear regression. That is
\begin{equation*}
x_{t} 
:=
\argmin_{a\in\R^d} \sum_{i=1}^I \bigg(\sum_{j=1}^d a_j\,L_{j-1}(\Delta_i) - \sigma_{t,i}^{IV} \bigg)^2\,,
\end{equation*}
where $x_t:= (x_{t,1}, \ldots x_{t,d})$. 
This results in a time-series of coefficients $(\{x_{t,j}\}_{j=1,\dots,d})_{t=1,\dots,T}$, which are shown in Figure \ref{fig:fda-coeffs}. That is our underlying stochastic process that the agent aims at combining has dimension five.

Using the time series $(\{x_{t,j}\}_{j=1,\dots,d})_{t=1,\dots,T}$, we train a neural-SDE model of the form
\begin{equation}\label{eq:neural-SDE}
    \d X_{t} = \mu(X_t)\,\dt+\sigma(X_t)\,\d W_t,
\end{equation}
with $\mu:\R^d\to\R^d$ and $\sigma:\R^d\to \S^d_{++}$,
by maximising the likelihood of the observed data subject to a probability integral transform (PIT) penalty ---  for further details on neural SDE estimation and why the PIT penalty decreases model misspecification, see \cite{choudhary2023funvol}. By telescoping the likelihood, the corresponding loss function for \eqref{eq:neural-SDE} is
\begin{equation*}
    -\sum_{t=1}^T
    \Bigg[ \frac{1}{2\Delta t}\left(\mu(x_t)\Delta t-\Delta x_t\right)^\intercal \Sigma^{-1}(x_t)
    \left(\mu(x_t)\Delta t-\Delta x_t\right)
    + \frac12 \log \det\big(\Sigma(x_t)\big)
    \Bigg]
    + \zeta\, \sum_{j=1}^d PIT_j,
\end{equation*}
where $\Delta x_t:= (\Delta x_{t,1}, \ldots, \Delta x_{t,d})$ with $\Delta x_{t,j}:=x_{t+1,j}-x_{t,j}$, $j  = 1,\ldots,d$, and
\begin{equation*}
    PIT_j := \int_0^1 \left(\sum_{t=1}^T \psi_h(u; z_{t,j}) -u\right)^2 \du,
    \quad \text{where} \quad
    z_{t,j} 
    :=
    \Phi\left(\frac{\mu_j(x_t)-\Delta x_{t,j}}{\sigma_j(x_t)}\right)\,,
\end{equation*}
and $\zeta$ is a hyper-parameter, $\Phi(\cdot)$ is the standard normal cumulative density function, $\psi_h(\cdot)$ is a kernel density with bandwidth $h$ (we use a Gaussian kernel with $h=0.01$), $\mu:= (\mu_1, \ldots, \mu_d)$,  $\Sigma=\sigma\sigma^\intercal$, and $\sigma_j=\sqrt{\Sigma_{jj}}$, $j=1,\dots,d$.

\begin{figure}
    \centering
    \includegraphics[width=0.7\textwidth]{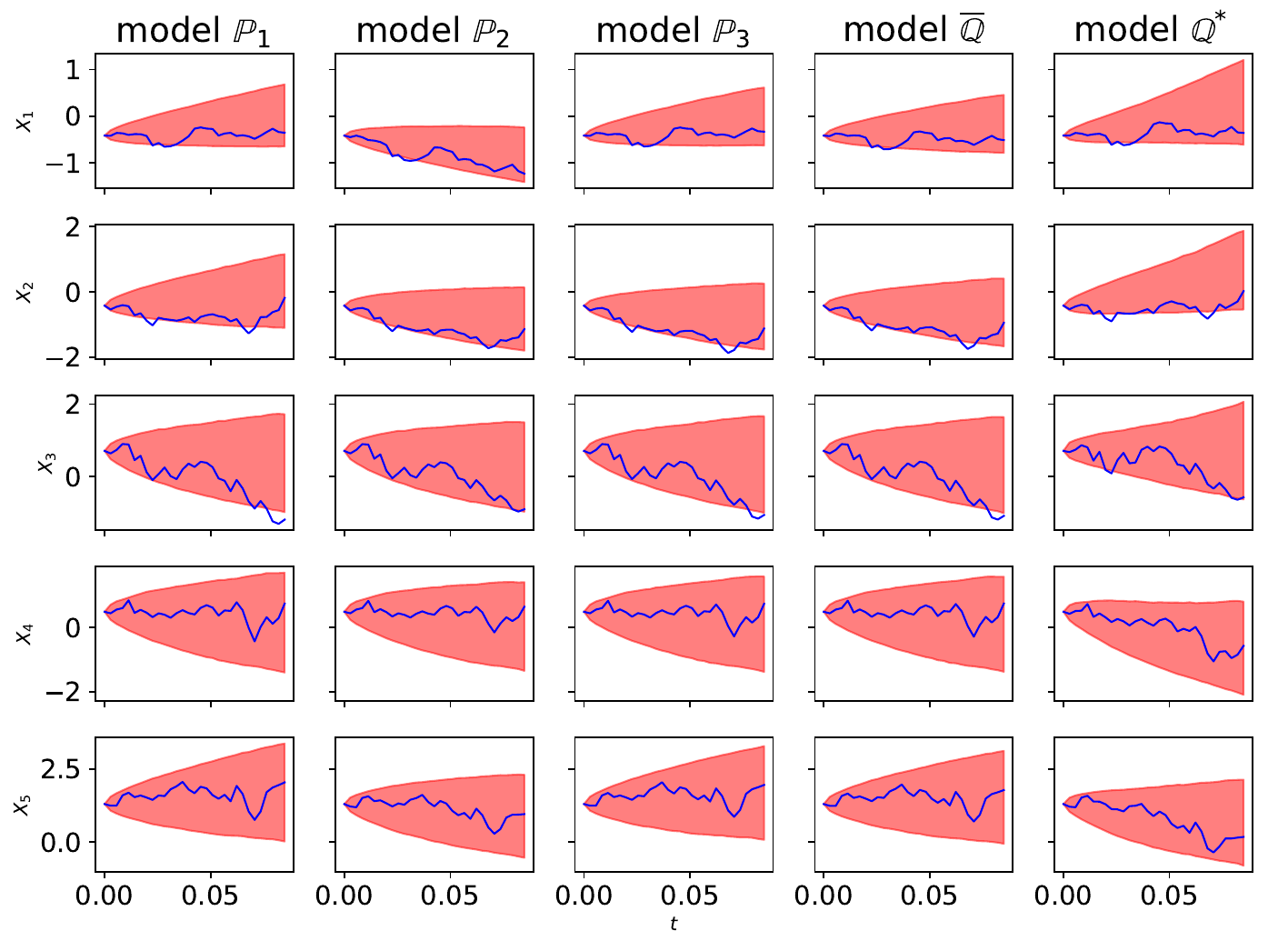}
    \caption{Simulations under the various models. The red bands show the 10\% and 90\% quantiles at each time point, while the blue line indicates a specific sample path.}
    \label{fig:smile-all-models}
\end{figure}
To train the neural-SDE, to estimate \eqref{eq:neural-SDE}, we employ two feed forward neural networks with five hidden layers and SiLU activation in all but the output layer for $\mu(x)$ and $\Sigma(x)$ separately. For the output layer, we use  a $10\tanh(\cdot)$ activation function to bound the outputs to the range $[-10,10]$ --- this improves the speed of convergence of the loss minimisation and also ensures that the resulting dynamics have strong solutions to the associated SDE.
 For $\Sigma(x)$, we reshape the output of the neural network into a $5$-dimensional lower triangular matrix $U(x)$ and set $\Sigma(x) = U(x)\,(U(x))^\intercal + 10^{-3} I$, where $I$ is the $5\times5$ identity matrix, to ensure strict positive definiteness. As both the drift and covariance matrices are outputs of feed-forward neural networks, with smooth, bounded activation functions (SiLU and tanh), they are guaranteed to be bounded and Lipschitz. Consequently, \Cref{asm:strong=sol-SDE} and \Cref{asm:compatibility}, and the  conditions in \Cref{thm:value-beliefs} are satisfied, so that we may apply our optimal measure change results. Furthermore, while we do not impose any static arbitrage constraints on the generated implied volatility smiles, \cite{choudhary2023funvol} finds that generated smiles (in their case entire surfaces) have a lower percentage of static arbitrage than the raw data itself.

To obtain the three expert models, we retrain \eqref{eq:neural-SDE} with a neural SDE model using data restricted to one of $[0,\frac13 T]$, $(\frac13 T,\frac23T]$, and $(\frac23 T, T]$, but only retrain the drifts $\mu(x)$. Note that, in our problem formulation,  the experts' models are absolutely continuous with respect to each other, thus they need to share the same volatility $\Sigma(x)$. This results in three distinct (experts) models, whose drift are $\mu^{(1)}(x)$, $\mu^{(2)}(x)$, and $\mu^{(3)}(x)$, and they share the volatility $\Sigma(x)$.

These three expert models together with the weights $\pi=(0.2,0.2,0.6)$ are then used to approximate the barycentre model subject to the constraint $\E^{\Q}[\partial_\Delta \sigma_\tau^{IV}(\Delta)|_{\Delta=\frac12}]=0.05$ using the elicitability approach, see \Cref{algo:elicit}, over a time horizon of $[0,\tau]$ years (with $\tau=\frac1{12}$). This constraint imposes that on average, the at-the-money skewness (i.e., at $\Delta = \frac12)$ of the IV smile at terminal time is equal to $0.05$ --- which is slightly higher than the average at-the-money skewness of each of the expert's model. We further place a higher weight on the expert model that is trained with more recent data, and equal weights on the other two expert models. \Cref{fig:smile-all-models} displays the sample paths of the 5-dimensional coefficients of the Legendre polynomials under the three experts' models (estimated via the neural SDE), the average drift model, and the optimal model. Using the simulations of the coefficients of the Legendre polynomials under the different models, we then generate, under each model, sample paths of IV smiles for the time horizon $[0, \tau]$.

\begin{figure}[ht]
    \centering
    \includegraphics[width=0.9\textwidth]{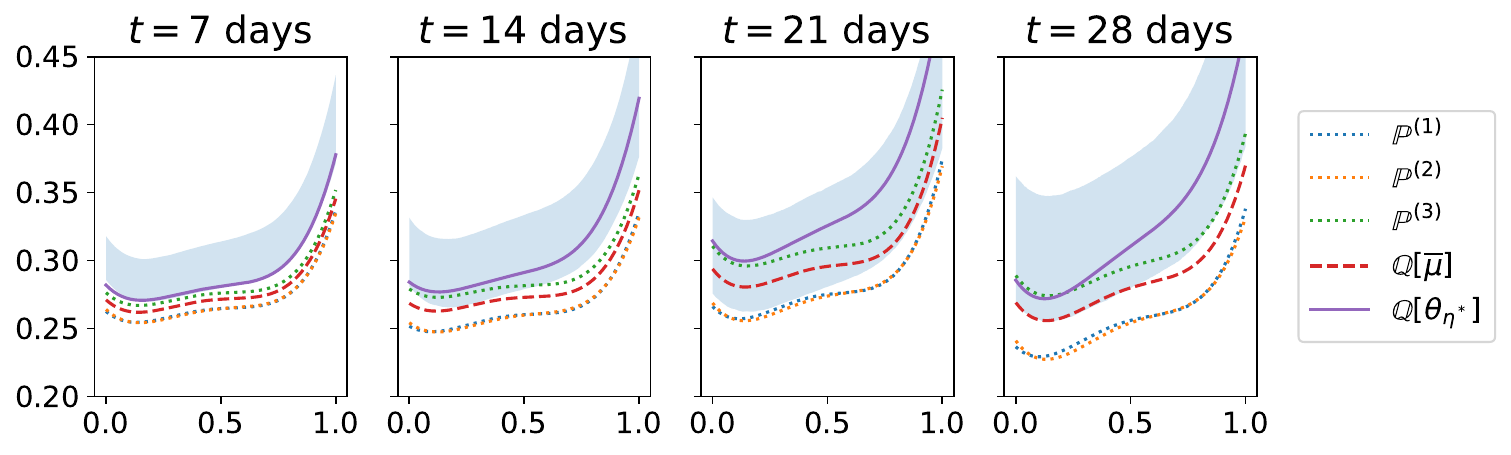}
   
    \caption{Simulated sample paths of IV smiles for times $t = 7, 14, 21, 28$ days under the experts models $\P^{(1)}, \P^{(2)}, \P^{(3)}$ (dotted), the average drift model $\Q[\bmu]$ (dashed), and $\Q[\theta_{\eta^*}]$ (solid). Shaded blue area are the 10\% to 90\% bands for the simulated IV smiles under $\Q[\theta_{\eta^*}]$.
    The weights for the experts are $\pi=(0.2,0.2,0.6)$.}
    \label{fig:IV-smiles-sim}
\end{figure}
\Cref{fig:IV-smiles-sim} shows a sample path of IV smiles from each expert, the model $\Q[\bmu]$, and the learnt constrained barycentre model $\Q[\theta_{\eta^*}]$, for four different times $t\in[0,28]$ days. The plots also display the $10\%$ to $90\%$ band of for $10,000$ sample paths from the $\Q[\theta_{\eta^*}]$ model. We observe that for later days, the bands increase and the models deviate from one another. Moreover, the displayed sample IV smile under $\Q[\theta_{\eta^*}]$ becomes steeper at $\Delta = \frac12$ to meet the constraint.

\section{Conclusions}
\label{sec:conclusions}

We present an approach to combine expert models, based on minimisation of the weighted KL divergence between the experts' models and a candidate model, that incorporates the agent's beliefs in the form of expectation constraints. The proposed methodology may be suitable to combine models in a multitude of settings in mathematical finance ranging from climate finance, to volatility modelling, and high-frequency trading. Moreover the resulting optimal model can be deployed as a generative model for, e.g., a dynamic risk-aware reinforcement learning approach to mitigating risks and optimising portfolio allocation \cite{coache2023conditionally,Jaimungal2023WP} or generating climate scenarios that feed into risk management for carbon planning \cite{hellmich2021carbon}.

\appendix
\clearpage

\bibliographystyle{siamplain}
\bibliography{main}

\end{document}

%% file: preamble.tex
\usepackage{todonotes}
\usepackage{dsfont}
\usepackage{booktabs}
\usepackage{wrapfig}
\usepackage{xcolor}

\usepackage{tikz}
\usetikzlibrary{decorations.markings, arrows.meta}
\renewcommand{\mathcolor}[2]{\color{#1}{#2}}
\definecolor{mgrn}{RGB}{0,255,100}
\tikzstyle{measure}=[shape=circle,draw=black,fill=blue!10,inner sep=0pt,minimum size=3pt] 

\newtheorem{assumption}{Assumption}[section]

\newcommand{\T}{{[0,T]}}

\newcommand{\R}{{\mathds{R}}}
\newcommand{\N}{{\mathds{N}}}

\newcommand{\Tr}{{\rm Tr}}

\newcommand{\thetaparam}{{\mathfrak{a}}}
\newcommand{\omegaparam}{{\mathfrak{o}}}
\newcommand{\tomega}{
{\tilde{\omega}}}

\renewcommand{\L}{{\mathcal{L}}}
\newcommand{\C}{{\mathcal{C}}}

\newcommand{\F}{{\mathcal{F}}}
\newcommand{\A}{{\mathcal{A}}}
\renewcommand{\S}{{\rm{S}}}

\newcommand{\mcK}{{\mathcal{K}}}

\newcommand{\mcQ}{{\mathcal{Q}}}
\newcommand{\mcH}{{\mathcal{H}}}

\newcommand{\mfM}{{\mathfrak{M}}}
\newcommand{\mfL}{{\mathfrak{L}}}

\newcommand{\mfa}{{\mathfrak{a}}}
\newcommand{\mfb}{{\mathfrak{b}}}
\newcommand{\mfc}{{\mathfrak{c}}}
\newcommand{\mfl}{{\mathfrak{l}}}
\newcommand{\mff}{{\mathfrak{f}}}
\newcommand{\mfg}{{\mathfrak{g}}}

\newcommand{\E}{{\mathbb{E}}}

\renewcommand{\P}{{\mathbb{P}}}
\newcommand{\Q}{{\mathbb{Q}}}
\newcommand{\Cov}{{\rm{Cov}}}
\newcommand{\Z}{{\mathbb{Z}}}

\newcommand{\dQ}{\d\Q}
\newcommand{\dP}{\d\P}

\newcommand{\oa}{{\overline{a}}}
\newcommand{\ob}{{\overline{b}}}

\newcommand{\ep}{\varepsilon}

\newcommand{\tT}{{t\in[0,T]}}

\newcommand{\Id}{{\mathds{1}}}

\newcommand{\bW}{{\overline{W}}}
\newcommand{\bmu}{{\bar\mu}}
\newcommand{\dmu}{{\Delta\mu}}
\newcommand{\blambda}{{\bar\lambda}}
\newcommand{\bgamma}{{\bar\gamma}}

\renewcommand{\d}{{\rm d}}
\newcommand{\dt}{{\rm d}t}
\newcommand{\du}{{\rm d}u}
\newcommand{\ds}{{\rm d}s}
\newcommand{\dW}{{\rm d}W}
\newcommand{\dbW}{{\rm d}\bW}

\newcommand{\hf}{{\hat{f}}}

\newcommand{\hg}{{\hat{g}}}

\newcommand{\msim}{{[m]}}
\newcommand{\bary}{{\mathfrak{b}}}

\DeclareMathOperator*{\argmin}{arg\,min}


%% file: ex_shared.tex
\usepackage{lipsum}
\usepackage{amsfonts, amsmath, amssymb}
\usepackage{graphicx}
\usepackage{epstopdf}
\usepackage[algo2e,linesnumbered,boxruled,lined,noend]{algorithm2e}
\SetAlFnt{\footnotesize}
\SetKw{Break}{break}
\SetKwComment{tcp}{$\triangleright$\ }{}

\ifpdf
  \DeclareGraphicsExtensions{.eps,.pdf,.png,.jpg}
\else
  \DeclareGraphicsExtensions{.eps}
\fi

\usepackage{enumitem}
\setlist[enumerate]{leftmargin=.5in}
\setlist[itemize]{leftmargin=.5in}

\newsiamremark{remark}{Remark}
\newsiamremark{hypothesis}{Hypothesis}
\crefname{hypothesis}{Hypothesis}{Hypotheses}
\newsiamthm{claim}{Claim}

\headers{KL Barycentre of Stochastic Processes}{Jaimungal and Pesenti}

\title{Kullback-Leibler Barycentre of Stochastic Processes\thanks{This version: April 9, 2025. First version: August 6, 2024.\\
Earlier version have been presented at the 2024 Oxford Man Institute (OMI) Machine Learning in Quantitative Finance conference (Oxford), the University of California Los Angeles, the University of California Santa Barbara, the University of Laval, Simon Fraser University, the University of Sydney, Banff IRS workshop on Modelling, Learning, and Understanding, and the Winter Workshop on Operations Research, Finance and Mathematics 2024 in Hokkaido.
\funding{SJ and SP would like acknowledge support from the Natural Sciences and Engineering Research Council of Canada (grants RGPIN-2024-04317, ALLRP 550308-20, and DGECR-2020-00333, RGPIN-2020-04289).}}}

\author{Sebastian Jaimungal\thanks{Department of Statistical Sciences, University of Toronto
  (\email{sebastian.jaimungal@utoronto.ca}, \email{silvana.pesenti@utoronto.ca}).}
\and 
Silvana M. Pesenti\footnotemark[2]}

\usepackage{amsopn}

\makeatletter
\newcommand*{\addFileDependency}[1]{
  \typeout{(#1)}
  \@addtofilelist{#1}
  \IfFileExists{#1}{}{\typeout{No file #1.}}
}
\makeatother

